    \documentstyle[12pt]{article}
    \def\l{\lambda}
    \def\m{\mu}
    \def\a{\alpha}
    \def\b{\beta}
    \def\o{\omega}
\begin{document}
    \baselineskip=22pt plus 1pt minus 1pt
  \centerline{\Large \bf Broken SU(3) symmetry in deformed
    even-even nuclei}
  \begin{center}
{\large N. Minkov$^*$\footnote[1]{e-mail: nminkov@inrne.acad.bg}, 
S. B. Drenska$^*$\footnote[2]{e-mail: sdren@inrne.acad.bg}, 
P. P. Raychev$^{*\#}$\footnote[3]{e-mail: 
raychev@bgcict.acad.bg}, 
R. P. Roussev$^*$\footnote[4]{e-mail: rousev@inrne.acad.bg} 
and Dennis Bonatsos$^\dagger$\footnote[5]{e-mail: 
bonat@cyclades.nrcps.ariadne-t.gr}\\
\medskip
$^*$    Institute for Nuclear Research and Nuclear Energy, \\
    72 Tzarigrad Road, 1784 Sofia, Bulgaria\\
\medskip
$^\#$  Dipartimento di Scienze Fisiche, Universit\`a di Napoli 
``Federico II'',\\
Mostra d' Oltremare Pad. 19, I-80185, Napoli, Italy\\
\medskip
 $^\dagger$   Institute of Nuclear Physics, N.C.S.R. ``Demokritos'',\\
    GR-15310 Aghia Paraskevi, Attiki, Greece}

\end{center}
\bigskip\bigskip
    \begin{abstract}

    A collective vector-boson model with broken SU(3)
    symmetry, in which the ground state band and the lowest $\gamma$ band 
belong to the same irreducible representation but are non-degenerate, is 
applied to several deformed even--even nuclei.  The model description of ground
    and $\gamma$ bands together with the corresponding B(E2) transition
    probabilities is investigated within a broad range of SU(3) irreducible
    representations $(\l ,\m )$.  The calculations show that the $(\l ,\m )$
    characteristics of rotational nuclei depend to a great extent on the
    magnitude of the SU(3) splitting between the ground and $\gamma$ bands.  
It is found that for weakly split
    spectra, the ground--$\gamma$ band coupling scheme is realized relevantly
    within narrow regions of ``favored'' $(\l ,\m )$ multiplets, while in the
    cases of strong splitting a description in which the ground 
band is situated alone in an irreducible representation is favored. 
 The obtained results are analyzed in terms of the bandmixing interactions.
    The possibility for a transition between the different collective SU(3)
    schemes is discussed.

    \end{abstract}
    \bigskip\bigskip
PACS Numbers: 21.60.Fw, 21.60.Ev

\newpage
    \section{Introduction}
    \label{sec-intr}

    The SU(3) symmetry group, which was introduced initially in nuclear
    theory as the symmetry group of $s,d$--shell nuclei \cite{Ell}, has also 
    been given meaning in the framework of the Dynamical Symmetry (DS) concept
 \cite{Dash,Wea,p:sp6r,Af}.  Based on the DS concept, it has been supposed that
    the SU(3) symmetry is inherent for the well deformed even--even nuclei, so
    that the low--lying ($L\leq 10$) collective states of these nuclei could
    be united into one or several SU(3) multiplets, labeled by the irreducible
    representations (irreps) ($\l ,\m$) of the group SU(3) \cite{p:descr}.
    The collective rotational Hamiltonian reduces this symmetry to the
    rotational group O(3) and thus the energy spectrum of the nucleus is
    generated.  In particular, it has been shown that in the rare earth
    nuclei the ground state band (gsb) and the first $\gamma$--excited band
    can be united into one split $({\l},2)$ multiplet appearing in a
    collective vector--boson scheme with broken SU(3) symmetry \cite{p:descr}.
 This scheme gives a satisfactory description of the energy levels and of the
    B(E2) transition ratios within and between the bands.  The success of the
    SU(3) scheme has inspired the extension of the concept of DS 
   in nuclei to the noncompact
    group Sp(6,$\Re$) \cite{p:sp6r,Rose1,Rose2,Rose3,Rose4,Fill}, which
    contains SU(3) as a maximal compact subgroup.  Alternatively, boson and
    fermion realizations of dynamical symmetries  have been used in the 
Interacting Boson Model (IBM) (having an overall U(6) symmetry) 
\cite{Arima1,Arima2,Arima3,Arima4} and the Fermion
    Dynamical Symmetry Model (with Sp(6,$\Re$)$\times$SU(2) and
    SO(8)$\times$SU(2) overall symmetries) \cite{FDSM1,FDSM2},
    respectively.  In spite of the different realizations these extended
    algebraic schemes in the appropriate limit include SU(3) as a DS group 
    which can be associated with the
    rotational limit of nuclear collective motion.

    Various model realizations of a broken SU(3) symmetry have been applied to
    the nuclei of the rare earth and actinide regions by using appropriately
    selected SU(3) irreps.  A microscopically justified one is the pseudo
$\widetilde{SU(3)}$ 
model (having an SU(3) abstract symmetry), in
    which the SU(3) irrep $({\l},{\m})$ used for a given nucleus depends
    on the filling of the Nilsson pseudo oscillator levels \cite{Jerry1}.  An
    alternative prescription for fixing the SU(3) quantum numbers $\l$
    and $\m$ is used in \cite{Ashe1,Ashe2} and is based on the original
    Elliott model \cite{Ell}.  In fact, the two schemes involve different SU(3)
    irreps for one and the same nucleus, indicating that with respect to the
    abstract SU(3) symmetry (beyond the particular realization), the choice of
    an adequate $({\l},{\m})$ multiplet for the given nucleus is 
    not unique.  The above circumstance naturally leads to the question
   of  whether the theoretically determined SU(3) irrep provides the best model
    description of the spectrum and how the pattern changes with varying
 $\l$ and $\m$.  It is therefore of interest to understand whether the
    appropriate irreps can be established directly on the basis of the
    available experimental data and whether they reflect the respective
    systematic behavior of the ground and $\gamma$ band rotational structure
    of deformed nuclei.  

In order to clarify these questions one should include
    in the study a large variety of $({\l},{\m})$ multiplets
and try to determine the ones favored by comparison to the experimental data. 
  Such an approach can be naturally applied in
    the framework of the vector--boson model scheme
    \cite{p:descr,a:over,p:matr}, in which the possible SU(3) multiplets are
    not  restricted by the underlying theory.   
This suggests that the SU(3) quantum numbers
    $\l$ and $\m$ are external parameters of the model scheme, allowing
    one to vary them so as to obtain the SU(3) irreps in which the
    experimental energies and transition probabilities are reproduced most
    accurately.  Once  such ``favored'' SU(3) irreps are found, one can apply
    them to the analysis of the collective dynamical characteristics of
    nuclei as well as to the discussion 
    of the physical meaning of the vector--boson scheme.

    An important characteristic of the SU(3) multiplets is the energy
    splitting of the even angular momentum states into the respective states
    belonging to the gsb and the $\gamma$--band.  The splitting is due to the
    reduction of the SU(3) symmetry in the nucleus and characterizes the
    mutual disposition of the two
 rotational bands within the multiplet.  Thus  one
    could expect that the possible existence of favored SU(3) irreps will
    depend on the energy splitting as well as on the intrinsic rotational
    structure of the bands.

    In this paper we report a global study of the broken SU(3) symmetry in
    deformed even--even nuclei, implemented through the use of the
    vector--boson formalism \cite{p:descr,a:over,p:matr}.  Motivated by the
    above considerations, we suppose that for a given rotational nucleus the
    physically significant features of this symmetry should be sought in
    certain regions of SU(3) irreps instead of a single fixed irrep. The
    aims of the work are concentrated on the following items:

    i) To study whether in the framework of the vector--boson scheme the
    available experimental information on the energy levels and transition
    probabilities could be used to estimate the SU(3) symmetry
    characteristics of the nucleus, in particular to outline the physically
    favored regions in the ($\l ,\m $) plane.

    ii) To study how the picture changes in the various nuclei, where
    different energy splittings between the ground state band and the 
first $\gamma$-excited band are observed, and if the SU(3) nuclei could be
    systematized accordingly.

    iii) To investigate the principal limits of applicability of the SU(3)
    symmetry in nuclei by analyzing the band mixing interactions in terms of
    the vector--boson formalism.

    We have considered eight rare earth nuclei ($^{164}$Dy, $^{164-168}$Er,
    $^{168,172}$Yb, $^{176,178}$Hf) and one actinide nucleus ($^{238}$U) 
  for which the
    model descriptions of the gsb and $\gamma$--band energy levels and the
    concomitant B(E2) transition ratios have been evaluated (in the form of
 root mean square fits) in SU(3) irreps within the range $10\leq\l\leq
    160$ and  $2\leq\m\leq 8$.  These nuclei represent regions of SU(3)
    spectra with different magnitudes of energy splitting between the gsb and
the first $\gamma$-band.   Though some other
    nuclei could also be included in the study, we shall see that the
    considered ones are sufficient to trace the most important features of
    SU(3) DS in collective rotational regions.

A few comments and clarifications are in place at this point:

i) The vector-bosons used in the vector-boson model \cite{p:descr}
do not possess any underlying 
physical content, in contrast to the bosons used in the Interacting Boson 
Model (IBM) \cite{IA}, which are understood as correlated fermion pairs
(see \cite{Bon} and references therein). The vector-bosons are the 
building blocks of the vector-boson model and the broken SU(3) symmetry, 
which do have a physical content, as it will be seen later. There is no
contradiction between the last two statements. The situation is similar to
that of the Schwinger boson realization of SU(2) 
\cite{BieLou,Bie,Mac}: The bosons
used for the realization do not bear any particular physical content 
themselves, but the SU(2) operators built out of them are the physically 
meaningful angular momentum operators.  

ii) The SU(3) symmetry discussed in this paper is a broken SU(3) symmetry,
in which the ground state band and the lowest $\gamma$ band belong to the 
same irrep but are non-degenerate. The lowest $\beta$-band is not contained
in the same irrep. The situation differs drastically from that of the SU(3) 
limit of IBM \cite{Arima3}, in which a pure SU(3) symmetry is the starting 
point, the ground state band sitting alone in an irrep, with the 
lowest $\gamma$- and $\beta$-bands belonging to the next irrep and being 
degenerate. The degeneracy of the even angular momentum levels of the 
lowest $\beta$- and $\gamma$-bands is a hallmark of the SU(3) symmetry of
IBM. 

   In  sec.~\ref{sec-mod} the vector--boson scheme, which in the
    lowest SU(3) irreps $(\l ,2)$ allows one to derive analytical expressions
    for the energy levels and transition probabilities \cite{p:descr}, is
    extended for calculations in the higher irreps with $\m >2$.  In
    sec.~\ref{sec-calc} we describe the numerical procedure and estimate the
    significance of the Hamiltonian parameters for the model description.  The
    obtained results and the corresponding theoretical analysis are presented
 in sec.~\ref{sec-res} while in sec.~\ref{sec-conc} the conclusions are given.

    \section{The Vector--Boson Model with a Broken SU(3) Symmetry}
    \label{sec-mod}

    \subsection{\it Basis and Hamiltonian}
    \label{subsec-basis}

    The present realization of the SU(3) dynamical symmetry is founded on the
assumption that the low lying collective states of the nuclear system can be
    constructed effectively with the use of two distinct kinds of vector
    bosons, whose creation operators $\mbox{\boldmath $\xi^{+}$}$ and
    $\mbox{\boldmath $\eta^{+}$}$ are O(3) vectors and in addition transform
    according to two independent SU(3) irreps of the type $({\l},{\m})=(1,0)$.
    The vector bosons are interpreted
    as the quanta of the elementary collective excitations of the nucleus.
    The basic states corresponding to the reduction chain
    \begin{equation}
    \label{eq:decomp}
    SU(3)\supset O(3)\supset O(2)
    \end{equation}
    can be constructed as polynomials in the vectors $\xi^{+}_{\nu}$ and
    $\eta^{+}_{\nu}$ ($\nu =1,0,-1$) acting on the vacuum state. The set of
    these states, usually denoted as
    \begin{equation}
    \label{eq:bast}
    \left|\begin{array}{c}({\l},{\m})\\{\a},L,M\end{array}\right\rangle
   , \end{equation}
    is known as the basis of Bargmann--Moshinsky \cite{bm:bas,m:bas}.  Since
    the chain (\ref{eq:decomp}) is not canonical, i.e.  in a given SU(3) irrep
    $(\l ,\m)$ more than one O(3) irreps $(L,M)$ appear, an additional
    quantum number $\a$ is introduced in order to distinguish the states with
    equal angular momenta $L$.  The quantum number $\a$ is related to the
    Elliott quantum number $K$ as $\a =(\m -K)/2$ \cite{a:over}.  The basis
    vectors (\ref{eq:bast}) are not orthogonal with respect to $\a$ and could
    be orthonormalized by means of the Hilbert--Schmidt procedure
    \cite{a:over}.  For a given $L$, the quantum number $\a$ runs over all
    integers in the interval \cite{a:over,m:bas}
    \begin{equation}
    \max\{ 0,\frac{1}{2}({\m}-L)\}\leq{\a}\leq \min\{ \frac{1}{2}
    ({\m}-{\beta}),\frac{1}{2}(\l +\m -L-\beta )\}\ ,
    \label{eq:minmax}
    \end{equation}
    where
    \begin{equation}
    \beta = \left\{ \begin{array}{ll}0, & \mbox{$\l +\m -L$ even} \\
                                     1, & \mbox{$\l +\m -L$ odd}
    \end{array} \right.
    \nonumber
    \end{equation}
    The values $\{ \a_{j} \}_{j=1\div d_{L}}$ with $\a_{j}<\a_{j+1}$
    determined in (\ref{eq:minmax}) label the different bands in which the
    angular momentum $L$ appears and $d_{L}$ is the multiplicity of the O(3)
    irrep (L,M) in the decomposition (\ref{eq:decomp}).
    Thus in the case of the $(\l ,\m\geq 4)$ multiplet ($\l >\m$; $\l
    ,\m$  even) the number $\a_{d_{L}}$ labels the ground
    state band with $L=0$, 2, 4, \dots, $\l$; $\a_{d_{L}-1}$ labels the 
$\gamma$-band
    with $L=2$, 3, \dots, $\l +2$; $\a_{d_{L}-2}$ corresponds to a band with
    $L=4$, 5, \dots, $\l +4$, etc.
    In the case $(\l ,2)$ the above scheme provides only two
    bands, the gsb and the $\gamma$-band, labeled by the quantum numbers
    $\a_{2} =1$ and $\a_{1} =0$ respectively.

    The collective Hamiltonian of the vector--boson scheme is based on the
    experimentally supported view that in deformed even--even nuclei the
    nuclear effective interaction is dominated by the collective quadrupole
    mode.  Thus it is assumed that the basic collective properties of these
    nuclei are determined by their angular and quadrupole momenta, which are
    naturally incorporated within the framework of the SU(3) DS.  The
    effective SU(3)--symmetry breaking Hamiltonian which should be an O(3)
    invariant \cite{Jd:scal,RRS2943} is constructed by using three basic O(3) 
    scalars as follows \cite{p:matr}:
    \begin{equation}
    \label{eq:v}
    V=g_{1}L^{2}+g_{2}L\cdot Q\cdot L +g_{3}A^{+}A\ ,
    \end{equation}
    where $g_{1}$, $g_{2}$ and $g_{3}$ are the parameters of the model;
    $L$ and $Q$ are the angular momentum and quadrupole operators respectively
    in the vector--boson realization:
    \begin{equation}
    \label{eq:Lm}
    L_{m}=-\sqrt{2}\sum_{\mu ,\nu} C^{1m}_{1\mu 1\nu}(\xi_{\mu}^{+}\xi_{\nu}+
    \eta_{\mu}^{+}\eta_{\nu})\ ,\ \ \ \ m=0,\pm 1\ ;
    \end{equation}
    \begin{equation}
    \label{eq:Qk}
    Q_{k}=\sqrt{6}\sum_{\mu ,\nu} C^{2k}_{1\mu 1\nu}(\xi_{\mu}^{+}\xi_{\nu}+
    \eta_{\mu}^{+}\eta_{\nu})\ ,\ \ \ \ k=0,\pm 1,\pm 2 \ ,
    \end{equation}
 with $C^{LM}_{lmlm^{\prime}}$ denoting the Clebsch--Gordan coefficients; the
    term $A^{+}A$ introduced originally in \cite{Af:ham} is constructed by the
    operator
    \begin{equation}
    A^{+}=\mbox{\boldmath ${\xi^{+}}^{2}{\eta^{+}}^{2}$}-(\mbox{\boldmath
    ${\xi}^{+}\cdot\eta^{+}$})^{2}
    \end{equation}
    and its Hermitian conjugate $A$. The physical content of $A^{+}A$ is
    discussed in \cite{p:matr} by assuming that the vectors
    $\mbox{\boldmath $\xi^{+}$}$ and
    $\mbox{\boldmath $\eta^{+}$}$ form a ``pseudospin'' doublet. This allows
    one to label the SU(3) multiplets by the numbers $(N,T)$ ($N=0$, 1, 2
\dots; $\
    T=\frac{1}{2}N,$ $\/\frac{1}{2}N-1,$ $\/\frac{1}{2}N-2$ \dots), which are
 related
    to $(\l ,\m)$ as
    \begin{equation}
    \label{eq:NT}
    N=\l +2\m \ ;\ \ \  T=\l /2 \ .
    \end{equation}
    The number $N$ corresponds to the number of vector bosons (interpreted as
related to the number of excitation quanta in the nucleus)
and $T$ is the ``pseudospin'' of the system of
    $N$ vector bosons. It has been shown that in these terms the operator
    $A^{+}$ can be considered as a creation operator of four particles with
    $L=0$ and $T=0$. In this way the operator $A^{+}A$ has been interpreted as
  the number operator of ``$\alpha$-like'' configurations in nuclei.

    \subsection{\it Energies and B(E2) transition probabilities}

    The eigenstate of the effective Hamiltonian (\ref{eq:v})
    with given angular momentum $L$ and energy ${\omega}^{L}$ can be
    constructed from the highest-weight (hw) basis states (with $M=L$) as
    follows:
    \begin{equation}
    \left|\begin{array}{c}({\l},{\m})\\{\o^{L}},L,L\end{array}\right\rangle
     =\sum_{j=1}^{d_{L}}C_{\omega ,j}^{L}
    \left|\begin{array}{c}({\l},{\m})\\{\a_{j}},L,L\end{array}\right\rangle\ .
    \label{eq:constr}
    \end{equation}
    Then the standard problem for eigenfunctions and eigenvalues reduces to
    the following homogeneous set of equations (written in matrix
    form) for the coefficients $C_{\o ,j}^{L}$:
    \begin{equation}
    (V_{j,j'}-\o^{L}\delta_{j,j'})(C_{\o ,j'}^{L}) =0, \qquad 
    \ \ \ \ j,j'=1\div d_{L}\ ,
    \label{eq:homog}
    \end{equation}
    where
    $V_{j,j'}\equiv
    \left\langle\begin{array}{c}({\l},{\m})\\{\a}_{j},L,L
    \end{array} \right| V
    \left|\begin{array}{c}({\l},{\m})\\{\a}_{j'},L,L
    \end{array}\right\rangle$ are the matrix elements of the Hamiltonian
    (\ref{eq:v}) between the hw basis states
    and $(C_{\o ,j'}^{L})$ is a vector--column.
    The eigenvalues $\o_{i}^{L}$,
    $i=1\div d_{L}$ (with $\o_{i} <\o_{i+1}$)  are determined by
    \begin{equation}
    \det (V_{j,j'}-\o^{L}\delta_{j,j'})=0\ .
    \label{eq:det}
    \end{equation}
    In the low-dimensional cases with $\m =2,4$, where $d_{L}=2,3$,
    Eq.~(\ref{eq:det}) can be solved analytically \cite{p:descr}, while
    in the cases with $\m >4$ one should find $\o_{i}^{L}$ by numerical
    diagonalization of the matrix $(V_{j,j'})$.
    We remark that the interaction $V$ mixes only basis
    states with neighboring values of the quantum number $\a$ so that the
    matrix $(V_{j,j'})$ is tridiagonal. The analytical form of the
    matrix elements of the operators $L\cdot Q\cdot L$ and $A^{+}A$ is given
    in Table~\ref{tab:matel}. Since the
    basis of Bargmann-Moshinsky is non-orthogonal, the matrix $(V_{j,j'})$
    is not hermitian. This fact does not affect
   the obtaining of real eigenvalues when the model
    parameters $g_{1}$, $g_{2}$ and $g_{3}$ are real.
    After obtaining the eigenvalues $\o_{i}^{L}$, one is able to derive the
    corresponding coefficients $C_{i ,j}^{L}\equiv C_{\o_{i} ,j}^{L}$,
    $i,j=1\div d_{L}$. Below we show how this can be done easily even in
    the cases with large dimension. For a given eigenvalue $\o_{i}^{L}$ we
    introduce the coefficients:
    \begin{equation}
    h_{i,j}=C_{i ,j}^{L}/C_{i ,1}^{L}\ ,\ \ \ j=1\div d_{L}\ ,
    \label{eq:hij}
    \end{equation}
    with $h_{i,1}=1$. Thus the set (\ref{eq:homog}) is reduced to a 
non-homogeneous set of $d_{L}-1$ equations for the coefficients $h_{i,j}$,
    $j=2\div d_{L}$. Then using the tridiagonal form of the matrix
    $(V_{j,j'})$, we derive the solution of this set (for
    arbitrary $d_{L}$) in the following recursive form:
    \begin{equation}
    \label{eq:hj}
    h_{i,j}=-\{ V_{j-1,j-2}h_{i,j-2}+(V_{j-1,j-1}-\o_{i}^{L})h_{i,j-1}\} /
    V_{j-1,j} \ ,\ \ \ j=3\div d_{L}\ ,
    \end{equation}
    with
    \begin{equation}
    \label{eq:h2}
    h_{i,2}=-(V_{1,1}-\o_{i}^{L})/V_{1,2} \ .
    \end{equation}
    After obtaining the coefficients $h_{i,j}$ and using the
    orthonormalization of the eigenfunction (\ref{eq:constr})
    we find the first coefficient $C_{i ,1}^{L}$:
    \begin{equation}
    \label{eq:c1}
    C_{i ,1}^{L}=\left(2\sum_{j=1}^{d_{L}}\sum_{j'=1}^{j}
    h_{i,j}h_{i,j'}
    \left\langle\begin{array}{c}({\l},{\m})\\{\a}_{j},L,L\end{array} \right.
    \left|\begin{array}{c}({\l},{\m})\\{\a_{j'}},L,L\end{array}\right\rangle
    -\sum_{j=1}^{d_{L}}h_{i,j}^2
    \left\langle\begin{array}{c}({\l},{\m})\\{\a}_{j},L,L\end{array} \right.
    \left|\begin{array}{c}({\l},{\m})\\{\a_{j}},L,L\end{array}\right\rangle
    \right)^{-1/2} \ ,
    \end{equation}
    where the analytical form of the overlap integrals
    $\left\langle\begin{array}{c}({\l},{\m})\\{\a}_{j},L,L\end{array} \right.
    \left|\begin{array}{c}({\l},{\m})\\{\a_{j'}},L,L\end{array}\right\rangle$
    is given in \cite{a:over}. The remaining coefficients $C_{i ,j}^{L}$,
    $j=2\div d_{L}$
    are then determined through (\ref{eq:hij}).  In such a way, applying the
    above procedure for all eigenvalues $\o_{i}^{L}$, $i=1\div d_{L}$ we
    obtain the matrix $(C_{i,j}^{L})$ which transforms the
    space of the basis functions
    $\left|\begin{array}{c}({\l},{\m})\\{\a_{j}},L,L\end{array}\right\rangle$
    into the space of
    the physical states (with determined energies)
    $\left|\begin{array}{c}({\l},{\m})\\{\o^{L}},L,L\end{array}\right\rangle$.

    In order to obtain the B(E2) transition probabilities in a given
    multiplet $(\l ,\m)$ one can use the action of the operator $Q_{0}$
    (\ref{eq:Qk}) on the hw basis state
    \begin{equation}
    Q_{0}\left|\begin{array}{c}({\l},{\m})\\{\a},L,L
    \end{array}\right\rangle =
    \sum_{\stackrel{k=0,1,2}{s=0,\pm 1}}a_{s}^{k}
    \left|\begin{array}{c}({\l},{\m})\\{\a}+s,L+k,L
    \end{array}\right\rangle \ ,
    \end{equation}
    where the coefficients $a_{s}^{k}$ are given in \cite{p:matr}.
    Then the matrix elements of $Q_{0}$ between the states with determined
    energy values (\ref{eq:constr}) can be derived in the form:
    \begin{equation}
    \label{eq:q0mel}
    \left\langle\begin{array}{c}({\l},{\m})\\{\o}_{i'}^{L+k},L+k,L
    \end{array} \right| Q_{0}
    \left|\begin{array}{c}({\l},{\m})\\{\o}_{i}^{L},L,L
    \end{array}\right\rangle =
    \sum_{j=1}^{d_{L}}C_{i,j}^{L}
    \sum_{s=0,\pm 1}a_{s}^{k}R_{\a_{j} +s,i'}^{L+k} \ .
    \end{equation}
    where $i$, $i'$ and $k$ take the values $i=1\div d_{L}$;
    $i'=1\div d_{L+k}$ and $k=0,1,2$; the matrix $C^{L}$ is determined for the
    states with angular momentum $L$ by Eqs
    (\ref{eq:hij})--(\ref{eq:c1}) and the matrix $R^{L}$ is defined as
    $R^{L}=(C^{L})^{-1}$.
    The most general form of the B(E2) reduced transition probability
    with $\Delta L=k$ between the level corresponding to the eigenvalue
    ${\o}_{i}^{L}$ and the level corresponding to ${\o}_{i'}^{L+k}$ is:
    \begin{eqnarray}
    \label{eq:be2}
    B(E2;{\o}_{i}^{L}\rightarrow {\o}_{i'}^{L+k}) & = &
    \frac{1}{2L+1}
    \left(\begin{array}{ccc}
    L+k & 2 & L \\
    -L  & 0 & L \end{array}\right)^{-2} \nonumber \\
    & \times & \left|
    \left\langle\begin{array}{c}({\l},{\m})\\{\o}_{i'}^{L+k},L+k,L
    \end{array} \right| Q_{0}
    \left|\begin{array}{c}({\l},{\m})\\{\o}_{i}^{L},L,L
    \end{array}\right\rangle\right| ^{2} \ .
    \end{eqnarray}

    \section{Parameters and Numerical Calculations}
    \label{sec-calc}

    We have realized numerically the general model scheme, given in the
    previous section.  Thus in a particular $(\l ,\m )$ multiplet ($\l >\m$;
    $\l ,\m$ even) we diagonalize the matrix $(V_{j,j'})$ for the various
    angular momenta $L$.  The gsb and $\gamma$-band levels with even $L$ are
    then determined as $E_{g}(L) =\o_{1}^{L}-\o^{0}$ and
    $E_{\gamma}(L)=\o_{2}^{L}-\o^{0}$ respectively, where $\o_{1}^{L}$ and
    $\o_{2}^{L}$ are the lowest and the next larger Hamiltonian eigenvalues
    respectively, and $\o^{0}=g_{3}\m^{2}(\l+\m+1)^{2}$ is the zero level
    eigenvalue. The $\gamma$--band energies with odd $L$ are determined as
    $E_{\gamma}(L) =\o_{1}^{L}-\o^{0}$.

    By using (\ref{eq:be2}) for the obtained energy levels, we calculate the
    following B(E2) interband transition ratios:
    \begin{eqnarray}
    \label{eq:rat}
    R_{1}(L)=\frac{B(E2;L_{\gamma}\rightarrow L_{g})}
    {B(E2;L_{\gamma}\rightarrow (L-2)_{g})} \ , &\ \mbox{L even,}
    \nonumber \\
    R_{2}(L)=\frac{B(E2;L_{\gamma}\rightarrow (L+2)_{g})}
    {B(E2;L_{\gamma}\rightarrow L_{g})} \ , &\ \mbox{L even,}\\
    R_{3}(L)=\frac{B(E2;L_{\gamma}\rightarrow (L+1)_{g})}
    {B(E2;L_{\gamma}\rightarrow (L-1)_{g})} \ , &\ \mbox{L odd,}
    \nonumber
    \end{eqnarray}
    and the gsb intraband ratios:
    \begin{equation}
    \label{eq:intra}
    R_{4}(L)=\frac{B(E2;L_{g}\rightarrow (L-2)_{g})}
    {B(E2;(L-2)_{g}\rightarrow (L-4)_{g})}, 
    \end{equation}
    where the indices $g$ and $\gamma$ label the gsb and the $\gamma$-band
    levels respectively. In the actinide nuclei the experimental
    information on the interband transitions does not suffice to provide any
    fits, so that in these cases (in particular in $^{238}$U) we consider only
    the intraband ratios (\ref{eq:intra}).

At this point it is important to estimate the significance of the Hamiltonian
    parameters $g_{1}$, $g_{2}$ and $g_{3}$ for the model calculations. The
    first parameter, $g_{1}$, applies only to the diagonal matrix elements of
    the Hamiltonian and contributes only to the rotational part of the energy
    levels.  The second and the third terms, $L\cdot Q\cdot L$ and $A^{+}A$,
    have diagonal as well as nondiagonal matrix elements (see
    Table~\ref{tab:matel}),
    so that the parameters $g_{2}$ and $g_{3}$ are significant
    for the rotational structure of the levels as well as for the band
    mixing interaction.  On the other hand, the diagonal contribution of the
    latter terms is responsible for the energy differences between the levels
    with equal angular momenta and different quantum numbers $\a$, which means
    that $g_{2}$ and $g_{3}$ are also significant for the splitting of the
    SU(3) multiplet.  


    In order to illustrate the above considerations, we refer to the 
particular case of the  $(\l ,2)$ irreps.  In a given ($\lambda  ,2$) 
irrep and for a given $L$ the general form of the Hamiltonian matrix
    elements is
    \begin{eqnarray}
    V_{i,j} &=& \langle \alpha_{i}|V|\alpha_{j}\rangle \nonumber \\
    &=& g_{1}\langle \alpha_{i}|L^{2}|\alpha_{j}\rangle+
        g_{2}\langle \alpha_{i}|L\cdot Q\cdot L|\alpha_{j}\rangle+
        g_{3}\langle \alpha_{i}|A^{+}A|\alpha_{j}\rangle \ ,
    \end{eqnarray}
    where the indices $i,j=1,2$ label the two $\alpha$- values: 
$\alpha_{1}
    =0$ and $\alpha_{2}=1$. Thus we have:
    \begin{eqnarray}
    V_{1,1} &=& \langle \alpha =0|V|\alpha=0\rangle, \\
    V_{2,2} &=& \langle \alpha =1|V|\alpha=1\rangle, \\
    V_{1,2} &=& \langle \alpha =0|V|\alpha=1\rangle, \\
    V_{2,1} &=& \langle \alpha =1|V|\alpha=0\rangle. 
    \end{eqnarray}
    Hence for the calculation of $V_{1,1}$ one needs from Table 1 
the values $\alpha =0,\ s=0$; for $V_{2,2}$ one needs $\alpha =1,\ 
s=0$; for $V_{1,2}$ one needs $\alpha =1,\ s=-1$; for $V_{2,1}$ one needs 
$\alpha =0,\ s=1$.

In this way one can easily see that in the case of $L$ being even 
(in which $\beta =0$ according to Eq. (4)) 
the diagonal terms of the Hamiltonian are 
(see
    Table~1)
    \begin{eqnarray}
\label{eq:V22}
V_{1,1} &=& g_{1}L(L+1)-g_{2}\left\{ (2\l 
+5)[L(L+1)-12]-6L(L-1)\right\},
\label{V11}    \\
    V_{2,2} &=& g_{1}L(L+1)-g_{2}\left\{ (2\l 
+5)L(L+1)-6L(L-1)\right\} \nonumber \\
            &+& g_{3}[4(\l +3)^{2}-2L(L+1)] \ ,
    \end{eqnarray}
    while the off-diagonal ones are
    \begin{eqnarray}
  \label{eq:V21}
    V_{1,2} &=& g_{2}12L(L-1), \label{V12}\\
    V_{2,1} &=& g_{2}6[-(2\l +5)+(2L+1)]+g_{3}2(\l +L+4)(\l -L+2) \ .
    \end{eqnarray}
In the case of odd $L$ (in which $\beta=1$ according to Eq. (4))
one finds (see Table~1)
\begin{equation}
V_{1,1} = g_1 L(L+1) -g_2 (2\l +5) [L(L+1)-12].
\end{equation}
    The gsb and $\gamma$--band energy levels are then obtained in the 
form
    \begin{eqnarray}
    E_{g}(L) &=& AL(L+1)-B\left( 
\sqrt{[1+CL(L+1)]^{2}+Df(L)}-1\right),
    \label{Eground} \\
    E_{\gamma}(L^{even}) &=& 2B+AL(L+1)+B
    \left(\sqrt{[1+CL(L+1)]^{2}+Df(L)}-1\right), \label{Egamm} \\
    E_{\gamma}(L^{odd}) &=& 2B+AL(L+1) \ ,
    \end{eqnarray}
    where
    \begin{eqnarray}
    A &=& g_{1}-(2\l +5)g_{2}-g_{3}, \label{termAAA}\\
    B &=& 6(2\l +5)g_{2}-2(\l +3)^{2} g_{3},  \label{termA}\\
    C &=& \frac{1}{6(2\l +5)}\frac{g_{3}}{g_{2}}, \\
    D &=& \frac{12}{B^{2}}[3g_{2}^2-g_{2}g_{3}],
    \end{eqnarray}
    and
    \begin{equation}
    f(L)=L(L-1)(L+1)(L+2) \ .
    \label{fterm}
    \end{equation}
These levels have been obtained in respect to the zero level eigenvalue
$ \omega^0= 4 g_3 (\l +3)^2$, as explained in the beginning of Section 3. 

    The linear combination of parameters $A$ could be interpreted 
as the
    inertia term, corresponding to the non-mixed part of the energy 
levels. The
    quantity $2B$ has the meaning of the $\gamma$--band bandhead, while 
$C$ and
    $D$ contribute to the mixed part of the energy levels. Note that 
$f(L)$
    coincides with the square of the $\Delta K=2$ bandmixing term of the
    Bohr-Mottelson model \cite{BM}.

    The above expressions indicate two specific features of the 
present model
    in the $(\l ,2)$-case:

    i) The odd $\gamma$-band levels, which in this case are  not 
mixed with any other levels, exhibit a rigid rotor behavior.

    ii) In the particular case $g_{3}/g_{2}=3$ the quantity $D$ 
vanishes, so that despite the splitting both the gsb and the $\gamma$ 
band contain only terms which are powers of $L(L+1)$. 

    It is also useful to rewrite Eqs (\ref{Eground}) and 
(\ref{Egamm}) in the    form:
    \begin{equation}
    E_{\nu}(L)=\left(\frac{1}{2{\cal{J}}_{0}}+    
\frac{1}{2{\cal{J}}_{L}}\right)L(L+1)+\frac{1}{2{\cal{J}}_{L}}
    \frac{D}{C}f(L)+\frac{C}{2{\cal{J}}_{L}}L^{2}(L+1)^{2} \ ,
    \label{EVMI}
    \end{equation}
    where $L$ is even, $\nu=g,\gamma$ and
    \begin{eqnarray}
    {\cal{J}}_{0} &=& \frac{1}{2A} \ , \\
    {\cal{J}}_{L} &=& \frac{1}{BC}\left(1+\frac{1}{2B}\Delta 
E(L)\right) \ ,
    \end{eqnarray}
    with
    \begin{equation}
    \Delta E(L)=E_{\nu}(L)-\frac{1}{2{\cal{J}}_{0}}L(L+1) \ .
    \end{equation}
    The first term in Eq.~(\ref{EVMI}) corresponds to the energy of a 
nonrigid rotor, the moment of inertia of which is angular momentum  
dependent. 
This dependence is similar to the one occuring in the 
 Variable Moment of Inertia (VMI) model  \cite{VMI}.
    The other (higher order) terms also depend on the angular momentum through
    $\Delta E(L)$. In such a way Eq.~(\ref{EVMI}) indicates that the
    influence of the Hamiltonian parameters on the energy 
characteristics of
    the model is essentially nonlinear.


Now, regarding the transition probabilities, we consider
    the recursive Eqs (\ref{eq:hj}) and (\ref{eq:h2}). We remark that
    since $g_{1}$ enters only in the diagonal part of the Hamiltonian, the
    subtraction $(V_{j-1,j-1}-\o_{i}^{L})$ in Eq. (\ref{eq:hj})
    eliminates its contribution to the determination of the eigenfunctions and
    consequently of the transition probabilities. More precisely, the
    contribution of the diagonal matrix elements to the eigenvalues is not
    affected by the diagonalization procedure. Also one can deduce easily
    that the eigenvalues, as solutions of Eq. (\ref{eq:det}), should be
    homogeneous functions of the parameters $g_{2}$ and $g_{3}$, so that
    after dividing both the numerators and the denominators of Eqs
    (\ref{eq:hj}) and (\ref{eq:h2}) by $g_{2}$ (or $g_{3}$) one concludes
    that the wave function coefficients and the transition probabilities
    should depend only on the ratio $g_{3}/g_{2}$ (or $g_{2}/g_{3}$). Thus
    while the energy description requires appropriate values of all
    Hamiltonian parameters, the inclusion of the transition probabilities in
    the fitting procedure only fixes the ratio $g_{3}/g_{2}$ (or
    $g_{2}/g_{3}$). We also remark that if one sets $g_{3}$ (or $g_{2}$) equal
    to zero, which means to neglect the term $A^{+}A$ (or $L\cdot Q\cdot
    L$), the transition probabilities will obtain some constant 
(non-adjustable) values.  It follows that both symmetry breaking terms are
necessary for a reasonable description of the B(E2) transition probabilities
    within the present SU(3) scheme.

    For obtaining the model description in a given SU(3) irrep $(\l ,\m )$ we
    have adjusted the Hamiltonian parameters to the low-lying experimental
    gsb and $\gamma$-band energy levels (up to L=8--10) and to the
    available transition ratios between them.  This is implemented by using 
the $\chi^2$ minimization procedure based on the Direction Set (Powell's) 
Method (DSM) \cite{Numrec}. The quality of the energy fits is measured by 
    \begin{equation}
    \label{eq:se}
    \sigma_{E} =\sqrt{\frac{1}{n_{E}}\sum_{L,\nu }\left( E_{\nu}^{Th}(L)-
                E_{\nu}^{Exp}(L)\right) ^{2}}, 
    \end{equation}
which is the standard energy rms-deviation with $n_{E}$ being equal to the 
number of the levels used in the fit and  $\nu =g,\gamma$ labeling the gsb and
the $\gamma$-band levels respectively.  By analogy, the quality of the 
fit of the transition ratios is measured by
    \begin{equation}
    \label{eq:sb}
    \sigma_{B} =\sqrt{\frac{1}{n_{B}}\sum_{L,\tau}\left( R_{\tau}^{Th}(L)-
                R_{\tau}^{Exp}(L)\right) ^{2}},
    \end{equation}
which is the rms-deviation of the transition ratios of Eq. (\ref{eq:rat}),
with $n_{B}$ being the number of the ratios used in the fit and $\tau =1,2,3,4$
    labeling the different types of ratios defined in Eqs (\ref{eq:rat}) and
    (\ref{eq:intra}).
The experimental data on energy levels are taken from \cite{Sakai}. The data
    on electromagnetic transitions are taken as follows:
    $^{164}$Dy \cite{164Dy1,164Dy2,NDS164};
    $^{164}$Er \cite{NDS164,164Er1};
    $^{166}$Er \cite{166Er1,166Er2,NDS166};
    $^{168}$Er \cite{168Er1,168Er2,NDS168};
    $^{168}$Yb \cite{NDS168,168Yb1};
    $^{172}$Yb \cite{172Yb1,172Yb2,NDS172};
    $^{176}$Hf \cite{176Hf1,NDS176};
    $^{178}$Hf \cite{178Hf1,NDS178};
    $^{238}$U \cite{NDS238}.
In this method weight factors are used in order to account for the different 
orders of magnitude of the energy levels and the transition ratios, which 
are fitted simultaneously. 
The Direction Set (Powell's) Method (DSM) \cite{Numrec} used here does not 
involve any computation of the gradient of any function and is directly 
applicable to the numerical  realization of the present model.
In addition we have tested an alternative fitting procedure involving
    numerical derivation, in which the differences between the model
    predictions and the experimental data are minimized with the use of an
iterational procedure of the Gauss-Newton type (GN) \cite{Alex}.  
In this method the energy levels and the transition ratios are again fitted
simultaneously, but this time with equal (unit) weight factors.
In this way we have
    found that the independent application of both fitting procedures, DSM
    and GN, in a given SU(3) irrep $(\l ,\m )$ leads to the same
  values for the Hamiltonian parameters.  This 
   fact shows that the theoretical scheme developed in the
    previous section provides a numerically stable model description.  It
    follows that in the various SU(3) multiplets the differing
    accuracy of the model description should be due only to the particular
    SU(3)--symmetry properties of the considered nucleus.


 At this point we should mention  that the simultaneous fitting of 
energy levels
    and transition probabilities is advantageous for our analyses.  In 
order to
    estimate the significance of such a procedure we refer to the 
calculations
    carried out in the framework of the pseudo $\widetilde{SU(3)}$ 
model
    \cite{Jerry1}.  In Ref.  \cite{Jerry1} only the ground and 
$\gamma$--band
    energy levels are used in the fits.  The B(E2) transition 
probabilities
    are determined using the wavefunctions obtained from the energy 
diagonalizations.  As a result the energy levels and the gsb
    intraband transition probabilities of the nuclei $^{160-164}$Dy,
    $^{164-168}$Er, $^{166,168}$Yb, $^{232}$Th and $^{234-238}$U are 
described
    satisfactorily. However, the obtained interband transition 
probabilities
    (Tables 6 and 7 of Ref. \cite{Jerry1}) do not reproduce 
accurately the
    experimental data. For example in the case of $^{168}$Er the 
interband
    ratio $R_{1}(L)$ [Eq.~(\ref{eq:rat})] obtains the values 
$R_{1}(2)=1.43$;
    $R_{1}(4)=3.0$; $R_{1}(6)=3.72$, while the experimental data give
    $R_{1}(2)=1.78$; $R_{1}(4)=4.81$; $R_{1}(6)=10.6$ 
\cite{166Er1,166Er2},
    i.e. for $L\geq 4$
    the experimental $R_{1}(L)$--ratios are not reproduced. Below we 
shall see
    that  in our calculations (with simultaneous fitting of energy 
levels and
    transition probabilities) the same ratio for the same nucleus 
obtains the
    values $R_{1}(2)=1.81$; $R_{1}(4)=5.34$; $R_{1}(6)=10.31$,  which 
are in
    very good agreement with the experimental data. Simultaneous 
energy--B(E2)
    fits have in addition been used in the framework of the 
pseudo-symplectic model
    \cite{Trol}, the advantages of such a procedure becoming clear also in 
this case.
    In addition we remark that the interband transitions play an important
    role in our study, since (as will be commented below) they carry
    information about the coupling of the  gsb and $\gamma$ bands
 into one SU(3)    multiplet.


    In the end of this section we should mention that the restriction on the
    energy levels used in the fits to  angular momentum values up to
   $L=8-10$ is appropriate because
 below this limit almost all gsb and $\gamma$--band levels
    of the investigated nuclei are observed experimentally.  Such a
    restriction allows one to study the systematic behavior of the broken
    SU(3) symmetry in the various nuclei on the basis of the same angular
    momentum values.  Thus we ensure that in most of the considered nuclei the
    even--spin levels belonging to the gsb are described together with their
    $\gamma$--band counterparts.  The splitting of the even--spin states as
    well as the band mixing strenghts are then correctly taken into
    account.  An exception is the nucleus $^{238}$U for which we consider the
    gsb up to $L=18$ and the $\gamma$-band up to $L=5$, due to the lack of 
further data on the $\gamma$-band.

    \section{Results and Discussion}
    \label{sec-res}

\subsection{\it Nuclei with small SU(3) energy splitting} 

    We have grouped the nuclei under study according to  the magnitude of
 the   SU(3) energy splitting. As a measure of the splitting we use the ratio
    \begin{equation}
    \label{eq:split}
    \Delta E_{2}=(E_{2_{2}^{+}}-E_{2_{1}^{+}})/E_{2_{1}^{+}}, 
    \end{equation}
    where $E_{2_{1}^{+}}$ and $E_{2_{2}^{+}}$ are the experimentally measured
    $2^{+}$ energy levels, belonging to the gsb and the $\gamma$-band
    respectively.  In the rare earth region this ratio varies 
    within the limits $7\leq\Delta E_{2}\leq 18$, while in the actinides one
    observes values in the range $13\leq\Delta E_{2}\leq 25$. 

     We start with the
    nuclei in which a small band splitting ratio $\Delta E_{2}\sim
    8-10$ is observed.  The three Er--isotopes $^{164-168}$Er and the nuclei
    $^{164}$Dy and $^{168}$Yb are representatives of this group of nuclei.
    As a typical example let us consider the $^{168}$Er case, where $\Delta
    E_{2}=9.3$. For this nucleus the model calculations are implemented in
    the SU(3)--irreps within the range $10\leq\l\leq 90$ and $\m =2,4,6,8$. The
    results obtained for the description of the energy levels 
 are shown in Fig. 1, where
    the corresponding rms-factors $\sigma_{E}$ are plotted as a function of
    the quantum number $\l$.  One finds that in the $(\l ,2)$--irreps
    $\sigma_{E}$ exhibits a well pronounced minimum at $\l =20$ with
    $\sigma_{E} =3.2$ keV.  In the $(\l ,4)$--irreps the minimum is found at
    $\l =16$, with $\sigma_{E} =3.8$ keV, while in the $(\l ,6)$--multiplets
    it is obtained at $\l =14$, with $\sigma_{E} =5.8$ keV.  One also finds
    that in the $(\l ,8)$--multiplets $\sigma_{E}$ obtains almost constant
    values, $\sigma_{E}\sim 11-12$ keV, without the presence of any minimum.
    Thus Fig. 1 shows that for $^{168}$Er the model scheme provides a
    clearly outlined region of ``favored'' multiplets in the $(\l ,\m
    )$--plane, including $\l =14-20$ and  $\m =2,4,6$.  Outside this region
    $\sigma_{E}$ increases gradually with the increase of $\l$ and for  $\l
    >40$ it saturates towards the values obtained in the $(\l ,8)$--multiplets.
 It is    also clear that the best description of the 
  energy levels corresponds to the
    multiplet $(20,2)$,  which provides the absolute $\sigma_{E}$--minimum
    observed in the considered variety of $(\l ,\m)$--multiplets (see Table
    \ref{tab:favor}).  In addition 
 we see that with the increase of the quantum
    number $\m$ the corresponding $\sigma_{E}$--minima increase in value and
    are shifted to smaller $\l$--values.  Regarding the transition
    probabilities, we remark that the B(E2) ratios (Eqs
    (\ref{eq:rat}), (\ref{eq:intra})) are reproduced with almost equal accuracy
    in the whole variety of multiplets, where the rms factor $\sigma_{B}$
    changes within very narrow limits ($\sigma_{B}=0.25-0.3$).  Actually the
    differences in the $\sigma_{B}$--values obtained in the different
    multiplets are of the order of the experimental uncertainties.  This
    result is due to the fact  that in the present model scheme the
    B(E2) transition probabilities depend only on the ratio $g_{3}/g_{2}$,
    which can be adjusted almost equally well in the various irreps.  The same
    behavior of $\sigma_{B}$ is observed in all investigated nuclei.

 Consider now the parameter values obtained for the nucleus $^{168}$Er in
    the various irreps, plotted in Fig. 2 as a function of the
    quantum number $\l$.  One sees (Fig. 2(a)) that in the $(\l
    ,2)$--multiplets $g_{1}$
    obtains only positive values 
which increase gradually with the increase
    of $\l$ and saturate to $g_{1}\sim 10$ keV.  In the irreps with $\m
    =4,6,8$, $g_{1}$ starts with negative values ($g_{1}\sim -13$ keV in the
    irrep $(12,4)$; $g_{1}\sim -44$ keV in the irrep $(12,8)$), but with
    increasing $\l$ it goes to positive values and saturates towards the
    values obtained in the $(\l ,2)$--multiplets.  The parameters $g_{2}$ and
    $g_{3}$ obtain only negative values, as it is shown in Fig. 2(b),(c).
    One also finds that both parameters decrease in absolute value with
    increasing $\l$ and saturate towards zero.  

Two comments should be made at this point: 

i) The small $g_{2}$ and $g_{3}$ absolute values obtained in the
    large--$\l$ region, $\l >40$, do not reduce the respective contributions
    of the second and the third terms of the 
Hamiltonian to the energy levels, since
    the matrix elements of the operators $L\cdot Q\cdot L$ and $A^{+}A$
    increase in absolute value as $\l$ increases (see Table~\ref{tab:matel}).
   Thus  one
    should not consider either $L\cdot Q\cdot L$ or $A^{+}A$ as small
    perturbations to the collective rotational energy. 


ii) As a consequence  of i), the diagonal contributions of the terms
$L\cdot Q\cdot L$ and $A^+A$ may dominate in the rotational structure 
of
the energy levels. Therefore the coefficient  of the $L^2$ term, 
$g_1$,
should not be thought of as the usual inertial parameter. Actually, 
we have
already shown that in the $(\l ,2)$ case the inertial term is 
determined
as a linear combination of all of the Hamiltonian parameters [see
Eq.~(\ref{termAAA})].  This is why the negative values of $g_1$ (as in 
Fig.
2(a)~) should not be considered as a surprise. For example, in the 
multiplet
$(16,2)$ the set of parameters $\{ g_{1},g_{2},g_{3}\}=\{
-1.159,-0.321,-0.590 \}$ (given in Table~\ref{tab:favor} for the 
nucleus
${}^{164}$Dy) gives for the inertial term the value $A=11.3$~keV, 
which is
reasonable for nuclei in the rare earth region.


Furthermore in Fig. 2(d)
    the ratio $g_{3}/g_{2}$ is plotted as a function of $\l$. One finds that
    $g_{3}/g_{2}$ decreases with increasing  $\l$. The change of
    this ratio compensates for the fact that the
    $A^{+}A$  matrix elements increase more rapidly with increasing
    $\l$ than the matrix elements of the operator $L\cdot Q\cdot L$ (below
    we shall further discuss the $\l$--dependence of these
    matrix elements, see also Table~\ref{tab:matel}). In such a way the smooth
    behavior of $g_{1}$, $g_{2}$, $g_{3}$ and $g_{3}/g_{2}$ obtained in the 
$(\l,\m)$--plane indicates that the present model scheme allows a consistent
    renormalization of the Hamiltonian parameters for the different SU(3)
    irreps.  For that reason one obtains reasonable model descriptions even in
    the multiplets outside the favored region.

    Almost the same picture has been obtained in the other nuclei with small
    SU(3) energy splittings.  For each of them we found a 
clearly outlined region
    of favored multiplets as for $^{168}$Er.  Thus in the $^{166}$Er case the
    favored multiplets are located within the region $\l =12-16$ and
 $\m =2,4,6$,
    while the best model description is obtained in the irrep $(16,2)$ (see
    Fig. 3). For the nucleus $^{164}$Er the favored multiplets are found
    within the region $\l =14-18$ and $\m =2,4,6$ and the best description
    corresponds to the irrep $(18,2)$ (see Fig. 4). For the nuclei
    $^{164}$Dy and $^{168}$Yb the best model descriptions are established in
    the multiplets $(16,2)$ and $(20,2)$ respectively (see Fig. 5(a),(b)).
    The rms factors $\sigma_{E}$ and $\sigma_{B}$ and the corresponding values
of the parameters obtained in the ``best'' irreps are listed in
    Table~\ref{tab:favor}.  We remark that in these irreps very good agreement
    between theory and experiment is found.  Also, we should mention that for
all the nuclei considered the parameters of the Hamiltonian exhibit the same
    numerical behavior in the $(\l ,\m)$--plane as the one observed for
    $^{168}$Er.

As a typical example of results given by the broken SU(3) symmetry for nuclei
with small SU(3) energy splitting we give in Table 3 the energy levels and 
transition ratios calculated for the nuclei $^{164}$Dy,  $^{164-168}$Er,
and $^{168}$Yb  and compare them
to the corresponding experimental data. The parameter values corresponding
to these results are the ones given in Table 2. Very good agreement between 
theory and experiment is observed. 

    On the so far presented results the following comments apply:

    i) Although the considered SU(3) scheme allows an appropriate
    renormalization of the Hamiltonian parameters which leads to reasonable
    model descriptions in all $(\l ,\m )$--multiplets under study, the
    calculations for the nuclei $^{164-168}$Er clearly outline corresponding
    regions of favored multiplets, where the descriptions 
of the energy levels are obtained
    essentially better than in the other irreps.  Since these regions are
    determined on the basis of the experimental gsb and $\gamma$--band
    characteristics, the above result can be interpreted as a natural physical
    signature of the broken SU(3) symmetry in these nuclei.  

    ii) For the nuclei with small bandsplitting, the best model descriptions
    are obtained in the multiplets with $\m =2$.  Generally one finds (see
    Figs 1, 3, 4) that for a fixed quantum number $\l$ the $(\l
    ,2)$--irreps give better results than the ones with $\m >2$.  Note
    that while the $(\l ,2)$--multiplets include only two bands
    (the gsb and the $\gamma$--band), the higher SU(3) irreps
    with $\m =4,6,8,...$ predict the presence of additional higher rotational
    bands.  Thus, for example, the $(\l ,4)$--multiplets predict an additional
    rotational band built on a $4^{+}$--state, which in the considered
    Er isotopes should be observed in the energy region of $3-4$ MeV. Indeed
    in $^{164-168}$Er nuclei such $4^{+}$ states are observed experimentally,
    but their energies are measured near $2$ MeV \cite{Sood}, which excludes
    the possibility of describing them together with the gsb and the
    $\gamma$--bands within the present model scheme.  This fact
    indicates that in the considered nuclei the broken SU(3) symmetry is
    naturally revealed in the lowest ($(\l ,2)$) irreps, where besides the
    gsb's and the $\gamma$--bands, no other bands are predicted.  Hence the
    inclusion of other rotational bands should be implemented by an extension
    of the present model scheme to a more general DS group, such as
    Sp(6,$\Re$).

    iii) The obtained results can be discussed in terms of the relationship
    between the collective model shape parameters $\beta ,\gamma$ \cite{BM}
    and the SU(3) irrep labels $(\l ,\m )$ \cite{JPD93}
    \begin{eqnarray}
    \label{eq:bemap}
    \beta^{2} &\sim & [\l^{2}+\l\m +\m^{2} +3(\l +\m )+3], \\
    \gamma & = & \tan^{-1}[\sqrt{3}(\m +1)/(2\l +\m +3)],
    \label{eq:gamap}
    \end{eqnarray}
    where $\beta$ and $\gamma$ characterize the axial and the non-axial
    quadrupole deformations of the nucleus respectively.  Eqs
    (\ref{eq:bemap}) and (\ref{eq:gamap}) are derived by requiring a
    correspondence between the invariants of the triaxial rotor group
    T$_{5}\wedge$SO(3) and these of the group SU(3) (for more details see
    \cite{JPD93}).  We should remark that while in \cite{JPD93} the above
    relationship is considered in a microscopic (shell model) aspect (via $(\l
    ,\m )$), in the present studies it could be used on a phenomenological
    level.  Thus we are able to make some estimates for the nuclear quadrupole
    deformations in terms of the favored SU(3) irreps.  As an example consider
    the favored $(\l ,\m )$ region obtained for the nucleus $^{168}$Er (Fig.
    1). One finds that for the multiplets $(20,2)$, $(16,4)$ and $(14,6)$
    Eq. (\ref{eq:gamap}) gives $\gamma =6.6^{\circ}$, $\gamma
    =12.5^{\circ}$, $\gamma =18.1^{\circ}$ respectively. It is clear that the
    best model description (the multiplet $(20,2)$) corresponds to relatively
    small non axial deformation of the nucleus. Such estimates can be made for
    the irreps appearing in the alternative SU(3) models.  In the pseudo
$\widetilde{SU(3)}$
model \cite{Jerry1} and in its 
pseudo-symplectic extension \cite{Trol}, the SU(3) irrep used for the nucleus
    $^{168}$Er is $(30,8)$, while in Ref. \cite{Ashe2} the same nucleus is
    associated with the multiplet $(78,10)$.  We see that although these
    multiplets lie outside the empirically favored $(\l ,\m )$ region, the
    corresponding values of the angle $\gamma$ ($\gamma =12.4^{\circ}$ for
    $(30,8)$ and $\gamma =6.5^{\circ}$ for $(78,10)$) are very close to the
    ones for $(16,4)$ and $(20,2)$ respectively.  We have obtained similar
    estimates for the other nuclei considered.  In all cases we found that the
experimental information on the energy levels and the transition probabilities
    implicitly indicates the presence of small non-axial deformations.


\subsection{\it Nuclei with medium and large SU(3) energy splitting}

    Let us now turn to nuclei in which large band splitting ratios $\Delta
    E_{2}>14-15$ are observed.  The nuclei $^{172}$Yb, $^{176}$Hf and
    $^{238}$U are characterized by such large $\Delta E_{2}$ values.  As a
    typical example consider the $^{172}$Yb case where $\Delta E_{2}=17.6$.
    In Fig. 6 the rms factors $\sigma_{E}$ obtained for this nucleus are
    given for the $(\l ,\m )$--multiplets in the range $10\leq\l\leq 160$  
and $\m
    =2,4,6$.  Here, compared with the previously considered nuclei, we find an
    essentially different picture.  We see that  in the $(\l ,2)$--multiplets 
    the $\sigma_{E}$--factor, which starts with $29$ keV at $\l =12$,
 decreases with increasing $\l$ and further at $\l >80-90$ saturates gradually 
to a constant value $\sigma_{E}\sim 6.5$ keV without reaching
    any minimum.  In the higher irreps with $\m >2$, $\sigma_{E}$ exhibits
    almost the same $\l$--dependence and the $\sigma_{E}$--values obtained for
    $\l >80-90$ lie on the average $0.1$ keV above the ones obtained in the
    corresponding $(\l ,2)$--multiplets. It follows that in the large $\l$'s
    ($\l\sim 100$) all considered multiplets practically provide equally
    accurate model descriptions.  A similar picture is observed in the nuclei
    $^{176}$Hf (with $\Delta E_{2}=14.2$) and $^{238}U$ (with $\Delta
    E_{2}=22.6$).  This is illustrated in Fig. 7 for the $(\l ,2)$
    multiplets.  In $^{176}$Hf we found that for large $\l$--values ($\l
    >70-80$) $\sigma_{E}$ saturates to the value $\sigma_{E}\sim 14.8$ keV
    (see Fig. 7(a)) and in the nucleus $^{238}U$ (Fig. 7(b)) $\sigma_{E}$
    obtains the values $\sigma_{E}\sim 1.6$ keV (see also
    Table~\ref{tab:favor}). It is clear that in the nuclei with large
    bandsplitting the calculations indicate the presence of a 
wide lower limit of
    the quantum number $\l$ instead of a narrow region of favored multiplets.
 Therefore in these nuclei one could make only rough estimates of the nuclear
    collective characteristics. Thus taking into account that in  general $\l
    >60$ and using Eq. (\ref{eq:gamap}) one finds that the strongly
    splitted SU(3) spectra should correspond to small ($\gamma <2^{\circ}$)
    non-axial deformations.

   It is also interesting to consider the nucleus $^{178}$Hf in which one
    observes a transition value of the band splitting ratio $\Delta E_{2}=
    11.6$.  In Fig. 8 the $\sigma_{E}$-values obtained for this nucleus are
    plotted for the $(\l ,2)$--multiplets in the range $10\leq\l\leq 100$.
    One sees that $\sigma_{E}$, which starts with the value $\sigma_{E}\sim 24$
    keV at $\l =12$, decreases with increasing $\l$ and in the region
    $30\leq\l\leq 40$ obtains a slightly expressed minimum where
    $\sigma_{E}\sim 7$ keV.  Further on, $\sigma_{E}$ increases slowly
    with $\l$ and near $\l =100$ grows up to the value
    $\sigma_{E}\sim 8$ keV.  Such a result indicates that the global
    $(\l ,\m )$ characteristics of the broken SU(3) symmetry are changed
    gradually from the nuclei with small band splitting to the nuclei where
    the splitting is large.

As a typical example of results provided by the broken SU(3) symmetry for
nuclei with medium and large SU(3) energy splitting we give in Table 4 the
energy levels and transition ratios calculated for the nuclei $^{172}$Yb, 
$^{176-178}$Hf, and $^{238}$U and compare them to the corresponding 
experimental data. The parameter values corresponding to these results 
are the ones given in Table 2. Good agreement between theory and experiment
is observed. 

    The following overall picture of the vector--boson model description in
    deformed nuclei can now 
be drawn.  In the nuclei where the band splitting is
    small, $\Delta E_{2}\sim 8-10$ ($^{164}$Dy, $^{164-168}$Er, $^{168}$Yb),
    the best model descriptions are found in clearly outlined regions of
    favored $(\l ,\m)$--irreps with relatively small values of the quantum
    number $\l$ ($16\leq\l\leq 20$) as well as of the quantum number $\m$
    ($2\leq\m\leq 6$).  Further with the increase of the splitting energy, 
as in
    the case of the nucleus $^{178}$Hf (with $\Delta E_{2}=11.6$), the favored
    multiplets are shifted gradually to larger $\l$-values ($\l\sim 40$) with
    slightly expressed $\sigma_{E}$ minimum. In the nuclei where large band
    splitting is observed, $\Delta E_{2}\sim 14-22$ ($^{172}$Yb, $^{176}$Hf,
    $^{238}$U), the present theoretical scheme provides almost equally good
    model descriptions in all $(\l ,\m)$--multiplets with $\l >60$ up to $\l
    =160$ and $\m =6$.  The estimates of the shape parameters show that the
    increasing magnitude of SU(3) splitting indicates an increase in the axial
    ($\beta$, Eq. (\ref{eq:bemap})) and decrease in the non-axial
    ($\gamma$, Eq. (\ref{eq:gamap})) deformations of nuclei.

\subsection{\it Band-mixing interactions}

    The above picture can be analyzed in terms of the collective SU(3)
    Hamiltonian and the respective band-mixing interactions.  For this purpose
    we study the $\l$--dependence of the Hamiltonian matrix elements and
    estimate their contribution to the energy spectrum in the large $\l$
    limit.  (Since the physically significant values of the quantum number
    $\m$ do not exceed $\m =8-10$, the large $\m$ limit is of no practical
    interest.) 

Let us consider the case of the $(\l ,2)$--multiplets (without
    restriction on the higher irreps) where in the even--spin states the
    dimension of the Hamiltonian matrix is $d_{L}=2$.  From the analytical
expressions given in Table~\ref{tab:matel} and Eqs (\ref{eq:V22})-(30)
we can estimate the $\l$ dependence of the total contribution of the 
second and third terms of the Hamiltonian, $L\cdot Q\cdot L$ and $A^+ A$, 
in the diagonal and off-diagonal matrix elements. In the large $\l$ limit 
the diagonal matrix elements $V_{1,1}$ and $V_{2,2}$ increase in absolute 
value, with the increasing of $\l$, as $\l$ and $\l^2$ respectively. 
The lower off-diagonal matrix element $V_{2,1}$ increases as $\l^2$, 
while the upper one, $V_{1,2}$, does not depend on $\l$. Hence the total 
contribution of the diagonal matrix elements in the eigenvalue equation 
(\ref{eq:det}) increases as $\l^3$, while the total contribution of the
off-diagonal ones increases as $\l^2$. It follows then that in the 
large-$\l$ limit the relative contribution of the off-diagonal 
(band-mixing) matrix elements of the operators $L\cdot Q\cdot L$ and 
$A^+ A$ (compared to the diagonal ones) decreases as $1/\l$. We note that 
in the case of multiplets with $\m > 2$ this contribution decreases 
even more rapidly.  

The above estimates show that the increase in the quantum number $\l$ is
    connected with the corresponding decrease in the mixing interaction
    between the gsb and the $\gamma$--band within the framework of the SU(3)
    symmetry.  Hence for the nuclei with small band splitting ($^{164}$Dy,
    $^{164-168}$Er, $^{168}$Yb) the relatively small $\l$--values ($\l\sim
    16-20$) indicate that the gsb and the $\gamma$--bands are strongly mixed.
    In the nuclei with a large band splitting ($^{172}$Yb, $^{176}$Hf,
$^{238}$U) the large $\l$'s correspond to a weak interaction between the two
    bands.  This means that for the latter nuclei the rotational character of
    the gsb and the $\gamma$ bands should be better developed.  Indeed the
    case of the nucleus $^{238}$U with a very large splitting ratio $\Delta
    E_{2} =22.6$ and a well pronounced rotational structure of the gsb supports
    the above supposition.

    The obtained $(\l ,\m )$ characteristics of deformed nuclei allow one to
    gain a physical insight into the vector--boson realization of a broken
    SU(3) symmetry. To illustrate this, we refer to the number of vector
    bosons $N$ determined for a given $(\l ,\m )$ multiplet through
    Eq.~(\ref{eq:NT}). We see that our results give a possibility to
    estimate the number $N$ for the nuclei under study. Thus we find that in
    the cases of small bandsplitting the favored $(\l ,\m )$ regions imply
    relatively small vector--boson numbers $N\sim 20-30$, while for the
    strongly splitted SU(3) spectra one has $N\sim 80-100$. Then taking into
    account the $\l$--dependence of the Hamiltonian matrix elements one
    deduces that the increase of $N$ can be connected to the decrease in the
    band-mixing interaction. In these terms the large $\l$ limit
    ($\l\rightarrow\infty $) boils down to the limit $N\rightarrow\infty$,
    which corresponds to an asymptotical decrease of the band interaction to
    zero. Thereby the multiplet splits into distinct noninteracting rotational
    bands and the SU(3) symmetry gradually disappears. This situation is
    equivalent to the group contraction process in which the SU(3) algebra
    reduces to the algebra of T$_{5}\wedge$SO(3) \cite{JPD93}. In such a way
    the SU(3) symmetry goes to that of the rotator. Note that an analogous
    transition is inherent in the IBM \cite{Arima3} and corresponds to an
    infinite number of bosons.  However one should not make any analogy between
the $s$-- and $d$--bosons of the IBM and the vector bosons since the latter are
    introduced as quanta of elementary collective excitations and can not be
    treated as coupled nucleon pairs.

\subsection{\it Discussion}

    The so far presented results and analyses allow us to discuss the
    applicability and the limitations of the broken SU(3) symmetry in nuclei.
 In addition the relevance of the gsb--$\gamma$ band coupling scheme can be
    clarified in terms of the investigated SU(3) multiplets. First, consider
    the weakly splitted spectra. In these cases the established regions of
    favored $(\l ,\m )$ irreps suggest a cutoff in the gsb near $L=16-20$,
    which in general is in agreement with the experimental picture observed in
    rare earth nuclei. We note that since the present model is addressed to
    the low--lying spectra (below the backbending), one  should not try to
    discuss the higher energy levels (in our studies we consider the  gsb 
and not the  Yrast band). On the other hand the narrow limits of the favored
    regions suggest relatively well determined values of the shape
    characteristics $(\beta ,\gamma )$. These considerations indicate that for
the nuclei with $\Delta E_{2}\sim 8-10$, both the gsb and the $\gamma$ band are
    united into one SU(3) multiplet in a consistent way. In the strongly
    split spectra the situation is quite different. The lack of any 
upper limit
    for the quantum number $\l$ suggests the presence of high angular momenta
    $L\sim 60-80$ which are not reasonable in the low--spin regime of nuclear
    collective motion. For the same reason one could not obtain clear
    estimates for the nuclear shape parameters as in the cases of favored $(\l
    ,\m )$ regions. Furthermore the large $\l$--values correspond to
    excessively large (Pauli forbidden) axial deformations of nuclei (see
    Eq. (\ref{eq:bemap})).  The above facts show that for 
    nuclei with a large splitting ratio $\Delta E_{2} >14$ the gsb--$\gamma$
    band coupling scheme comes up against basic difficulties in the
    consistent treatment of nuclear collective characteristics. At least these
nuclei should be referred to the limiting case in which the two bands are 
weakly coupled in the framework of one SU(3) multiplet. This is a 
worth-mentioning 
    finding which could be interpreted as an indication for a possible
    rearrangement of the collective rotational bands in different SU(3) irreps.
    We can point out two experimental pieces of evidence supporting this
    supposition:

    i) For the nuclei with a large $2^{+}$ splitting the number of the
    experimentally observed gsb--$\gamma$ interband transitions is essentially
    smaller than the one in the nuclei where $\Delta E_{2}$ is small.
    Moreover in the nucleus $^{238}$U such transitions have not been observed.

    ii) Consider the mutual disposition of the second $2^{+}$ collective
    levels $E_{2_{2}^{+}}$ (the $\gamma$--band bandhead) and the corresponding
    second $0^{+}$ levels $E_{0_{2}^{+}}$ (the $\beta$--band bandhead) of
    rotational nuclei \cite{Sakai}. Note that for the nuclei with small
    $\Delta E_{2}$ ($^{164}$Dy, $^{164-168}$Er, $^{168}$Yb) one observes
    $E_{2_{2}^{+}}<E_{0_{2}^{+}}$ (for example, for $^{168}$Er one has 
    $E_{2_{2}^{+}}=0.821$ MeV and $E_{0_{2}^{+}}=1.217$ MeV). For the nucleus
    $^{178}$Hf, which has a transitional
 $\Delta E_{2}$--value, both energies are almost equal
    ($E_{2_{2}^{+}}=1.175$ MeV,  $E_{0_{2}^{+}}=1.199$ MeV).  For the nuclei
    with large $2^{+}$ splitting ($^{172}$Yb, $^{176}$Hf, $^{238}$U) one finds
    $E_{2_{2}^{+}}>E_{0_{2}^{+}}$ (for example, for $^{172}$Yb one has 
    $E_{2_{2}^{+}}=1.466$ MeV and  $E_{0_{2}^{+}}=1.042$ MeV). The latter
    observation indicates that in the nuclei with $\Delta E_{2}>14$ the gsb
    and the $\gamma$-band could be situated in distinct SU(3) multiplets.

    We remark that our analysis is consistent with the results obtained for
    the nucleus $^{238}$U in the framework of the pseudo
$\widetilde{SU(3)}$
and pseudo symplectic schemes
    \cite{Jerry1,Trol}. It is shown that the ``leading'' irrep appearing for
    this nucleus is $(54,0)$, which indicates that in this case the gsb
    probably belongs to a separate irrep.
    Actually, the obtained systematic properties of the SU(3) symmetry in
    deformed nuclei could be interpreted as the manifestation of a more
    general DS in nuclear collective motion. In this respect the gsb--$\gamma$
    band coupling schemes and the IBM collective scheme could be considered
    rather as complementary than as alternative schemes. The dynamical 
mechanism causing the rearrangement of rotational bands in the various 
SU(3) irreps could receive attention in the framework of a larger DS group. 


A more detailed comparison between the features of the present scheme and 
these of the Interacting Boson Model (IBM) is now in place. 
 As has already been mentioned, in IBM
    the lowest $\gamma$- and $\beta$- bands belong to the same SU(3) 
irrep
    $(2N-4,2)$, while the gsb remains alone in the most symmetric 
irrep
    $(2N,0)$ (where $N$ is the total number of active bosons). 
Formally both band coupling schemes, the gsb--$\gamma$ scheme (of the present 
model) and the $\beta$--$\gamma$ scheme (of IBM) could be referred to SU(3) 
multiplets of the type $(\l ,2)$. However, in the exact SU(3) limit of the 
original IBM-1 \cite{Arima3} the appearing $(\l ,2)$--multiplets are 
degenerate
    with respect to the Elliott quantum number K. This degeneracy (which is
    generally in disagreement with the experimental situation) can be removed
    in several ways. One possible way is to break the exact SU(3) symmetry.
    This can be achieved (see \cite{Bon} and references therein) 
by using in the usual IBM-1 Hamiltonian
  of the SU(3) limit
    \begin{equation}
    H_{SU(3)}=-\kappa (Q\cdot Q)+{\kappa}'(L\cdot L)
    \label{HIBM}
    \end{equation}
    the operator
    \begin{equation}
    Q_{m}=(d^{+}\tilde{s}+s^{+}\tilde{d})^{2}_{m}+\chi (d^{+}\otimes
    \tilde{d})^{2}_{m} \ ,
    \label{QIBM}
    \end{equation}
    where $\kappa$ and ${\kappa}'$ are the model parameters, and 
$s^{+},d^{+}$ $(s,d)$ are the creation (annihilation) operators for the $s$ 
and $d$ bosons, with $\tilde{s}=s$ and $\tilde{d}_{m}=(-1)^{m}d_{-m}$. 
In the case of $\chi =    -\sqrt{7}/2$, $Q$ is
    a generator of SU(3) and the exact SU(3) Hamiltonian is obtained. If $\chi
    =0$, $Q$ is a generator of O(6) and the Hamiltonian of Eq.~(\ref{HIBM}) is
 not an SU(3) Hamiltonian anymore. The case $-\sqrt{7}/2<Q<0$ corresponds to a 
broken    SU(3) symmetry. The $\beta$ and $\gamma$ bands then belong to one 
splitted $(\l ,2)$ multiplet. In such a way the $\beta$--$\gamma$ band 
coupling scheme of the IBM becomes very similar to the present gsb--$\gamma$ 
scheme.

    The same problem can also be solved by adding to $H_{SU(3)}$ some
    higher-order interaction terms. Such a term is the so-called O(3) 
scalar
    shift operator  which corresponds to a three-body interaction
    \cite{BMI85}. This operator, usually denoted by $\Omega$, possesses a 
realization in terms of $s$ and $d$ bosons and is equivalent
    to the second term of the vector-boson Hamiltonian 
[Eq.~(\ref{eq:v})]. It is not diagonal in the Elliott basis \cite{Ell},
 its eigenvalues in the $(\l 
,2)$
    irreps being \cite{BMI85}:
    \begin{eqnarray}
    \langle\Omega\rangle &=& \sqrt{6}{[L(L+1)-12]}(2\l +5) \ ,\ \ \ 
L={\rm odd} \\
    \langle\Omega\rangle &=& \sqrt{6}\{ (L-2)(L+3)(2\l +5) \nonumber 
\\
    &\pm & 6\sqrt{L(L+1)(L-1)(L+2)+(2\l+5)^{2}}\}\ ,\ \ \ L={\rm even} \ ,
    \label{omega}
    \end{eqnarray}
    with $\langle\Omega\rangle =0$ for $L=0$. The double sign in
    Eq.~(\ref{omega}) breaks the degeneracy between the levels of the
$\beta$ and
    $\gamma$ bands and thus the multiplet is splitted. Again we find 
that the
    situation is very similar to that of the present SU(3) symmetry 
model.
    Moreover if we consider the vector--boson Hamiltonian 
[Eq.~(\ref{eq:v})] with
    $g_{2}=1$ and $g_{3}=0$, the square root terms of Eqs. 
(\ref{Eground}) and
    (\ref{Egamm}) coincide exactly with the square root term in 
Eq.~(\ref{omega}).
    Thus in this case the $\beta$--$\gamma$ band coupling scheme of the 
IBM and the gsb--$\gamma$ scheme of the present model are characterized 
by the same analytical expression for the energy splitting:
    \begin{equation}
    |E_{\gamma}(L)-E_{\nu}(L)|\sim\sqrt{f(L)+(2\l +5)^{2}} \ ,\ \ \ 
L={\rm even} \ ,
    \end{equation}
    where $\nu$ labels the gsb (in the present model) or the 
$\beta$--band (in
    IBM), and $f(L)$ is defined in Eq.~(\ref{fterm}). Note that while in 
the
    gsb--$\gamma$ scheme the (+) sign
    in Eq.~(\ref{omega}) always corresponds to the $\gamma$--band and the 
($-$) sign always  corresponds to the gsb (i.e. the gsb levels are always 
below the  respective $\gamma$--band levels), in the $\beta$--$\gamma$ 
scheme the $\pm$ correspondence depends on the mutual displacement of the 
levels and  may be changed.

    A comment concerning the transition probabilities in the
    vector--boson model and in the IBM can be made here. While in the first
    model the $\gamma$--gsb interband E2 transitions are naturally
    incorporated, in the second one (in the exact SU(3) limit) they are
    forbidden. This type of transitions can be allowed in IBM by modifying the 
quadrupole transition operator similarly to Eq.~(\ref{QIBM}), i.e.  by breaking
 the exact SU(3) symmetry (see \cite{Bon} and references therein).

   It should also be mentioned that the large $\l$-values appearing in 
our work for the nuclei with large $\Delta E_{2}$ splitting correspond to 
the large $\l$-values obtained with the introduction of $g$-bosons in the 
framework of the $sdg$-IBM \cite{Wu1982,Akiyama}, where the
    band cutoffs are shifted towards higher angular momenta.

    The above considerations illustrate some differences between the present 
model and the IBM, as well as some common schematic features of both models.  
The present analysis also 
allows one to estimate the relative appropriateness of these model schemes  
for a particular  rotational nucleus or group of nuclei. Our results suggest 
that for nuclei with small $\Delta E_{2}$ splitting ratio the gsb--$\gamma$ 
band coupling scheme of the vector-boson model is more appropriate than the
    $\beta$--$\gamma$ scheme of IBM. As a typical example for this 
case we consider the nucleus ${}^{168}$Er, in which a large number of 
$\gamma$--gsb interband E2 transitions are observed 
\cite{166Er1,166Er2,NDS166}. For the nuclei with large gsb--$\gamma$ splitting
 the $\beta$-$\gamma$ coupling scheme of IBM seems to
    be more appropriate. As a typical example for this case we 
consider the nucleus ${}^{238}$U.

   In conclusion, the indicated rearrangement of the rotational bands in 
various SU(3) multiplets can be interpreted as an interplay between the 
different DS schemes of the vector-boson model and the IBM.  The dynamical 
mechanism causing this rearrangement should be considered in the
framework of  the DS of a group larger than SU(3).


    \section{Conclusions}
    \label{sec-conc}

    In this paper we have studied the broken SU(3) symmetry in deformed
    even--even nuclei via the formalism of the collective vector-boson model.
    We assume that the physically meaningful properties of this symmetry
    are developed in certain regions of $(\l ,\m)$ irreps, instead of one
    fixed irrep. In this way there is no microscopic input in the determination
of the $(\lambda, \mu)$ irrep of SU(3) suitable for each nucleus, the 
quantum numbers $\lambda$ and $\mu$ being treated as free parameters and 
fitted to the experimental data. 
The available experimental information on energy
    levels and transition probabilities allows one to identify two kinds of
nuclei with SU(3) symmetry:

    i) The nuclei with weak $2^{+}$ splitting ($\Delta E_{2} <12$, defined 
in Eq. (24)), for which we obtain narrow regions of favored SU(3) irreps (in 
general one has $16\leq\l\leq
    20$ and $2\leq\m\leq 6$).  In these regions the gsb--$\gamma$ band coupling
    scheme gives good model estimates of the nuclear collective
    characteristics under study.

    ii) The nuclei with strong $2^{+}$ splitting ($\Delta E_{2} >12$, 
defined in Eq. (24)), for
    which the successful model description requires large values of the
 quantum number $\l$ ($\l >60-80$) without any presence of particular regions
    of favored irreps.  In these nuclei the applied SU(3) scheme allows only
    rough estimates of nuclear collective properties. These nuclei are very 
good rotators, so that a pure SU(3) scheme, like the one of IBM, appears as
more appropriate.

    In such a way we find that the violation of the SU(3) symmetry, measured
    by the splitting ratio $\Delta E_{2}$ (defined in Eq. (24)), 
determines to a great extent the
    most important SU(3) properties of deformed nuclei.  

A  systematic analysis of the gsb--$\gamma$ band-mixing
    interaction on the basis of the collective vector--boson model leads to 
the following conclusions: Increasing number of vector bosons $N$ corresponds
 to the increase in the splitting of the multiplet and leads to decrease in
    the band-mixing interaction within the framework of the SU(3) symmetry.  In
    these terms the large $\l$ limit corresponds to $N\rightarrow\infty$ and
    has the meaning of SU(3) group contraction. In the limiting case the SU(3)
    symmetry is completely destroyed and the bands can not be united anymore
    in one SU(3) multiplet. Following the above analysis, we conclude that the
    strongly split spectra should be considered as special cases in which
    the gsb and the $\gamma$--bands are weakly coupled. Furthermore the
    experimental and theoretical examples given for these spectra indicate the
    possibility for rearrangement of the two bands into distinct irreps. This
    finding suggests the presence of a transition from the gsb--$\gamma$ band
    coupling scheme (in the nuclei with small $\Delta E_{2}$) to an alternative
    collective scheme (in the cases of large $\Delta E_{2}$), in which the gsb
    is situated in a separate irrep. In other words the broken SU(3) scheme 
is favored in the case of weak $2^+$ splitting, while strong $2^+$ 
splitting favors SU(3) schemes like the one of the IBM, in which the gsb 
is situated in a separate irrep. 

The collective dynamical mechanism
    causing such a transition from the broken SU(3) of the present model to 
the pure SU(3) of the IBM  could be sought in the framework of the more
    general DS group Sp(6,$\Re$). In such a framework the lowest $\beta$-band,
absent from the broken SU(3) model considered here, could be included, 
belonging to an irrep different from the one in which the gsb and the lowest 
$\gamma$-band are located. These will be the subjects of a future 
investigation.

\bigskip
{\Large{\bf Acknowledgments}}
\medskip

The authors are thankful to S. Pittel for illuminating discussions and for a 
careful reading of the manuscript. 
    This work has been supported in part by the Bulgarian National Fund for
    Scientific Research under contracts no F--547 and no F--415.  One of the
    authors (DB) has been supported by the EU under contract ERBCHBGCT930467 
and by the Greek General Secretariat of Research and Technology under 
contract PENED95/1981. Another author (PPR) has been supported by the 
Istituto Nazionale di Fisica Nucleare (INFN) and the Italian Ministero
dell' Universit\`a e della Ricerca Scientifica e Tecnologica (MURST).

    \begin{table}
    \caption{Matrix elements of the operators $L\cdot Q\cdot L$ and
    $A^{+}A$ between the basis states of Eq. (\protect\ref{eq:bast}).}
    \bigskip
    {\small
    \begin{center}
    \begin{tabular}{lc}
    \rule{0em}{2.2ex}
    &  \\
    \hline
    \hline
    \vspace{1pt} \\
    $s$&  $
    \left\langle\begin{array}{c}({\l},{\m})\\{\a}+s,L,L
    \end{array} \right| L\cdot Q\cdot L
    \left|\begin{array}{c}({\l},{\m})\\{\a},L,L
    \end{array}\right\rangle $ \\
    \vspace{1pt} \\
    \hline
    \vspace{1pt} \\
     0 &  $
     \begin{array}{l}4{\a}[L(L+1)-3(L+2{\a}-{\m}+{\b})^{2}]-2({\l}+{\m}-L-{\b}
     -2{\a})[L(L+1) \\ -3({\m}-2{\a})^{2}]-(L-2{\m}+4{\a}+{\b})(2L+3)
     (L+1+3{\b})\end{array} $ \\
    &  \\
     1 &
     $-6({\l}+{\m}-L-{\b}-2{\a})({\m}-2{\a}-{\b})({\m}-2{\a}-{\b}-1) $\\
    &  \\
    $-1$ &
     $12{\a}(L+2{\a}-{\m})(L+2{\a}-{\m}-1)$ \\
    \vspace{1pt} \\
    \hline
    \hline
    \vspace{1pt}\\
    $s$&  $
    \left\langle\begin{array}{c}({\l},{\m})\\{\a}+s,L,L
    \end{array} \right| A^{+}A
    \left|\begin{array}{c}({\l},{\m})\\{\a},L,L
    \end{array}\right\rangle $ \\
    \vspace{1pt}\\
    \hline
    \vspace{1pt} \\
     0 &  $
     \begin{array}{l}-\frac{4}{3}{\a}\{ ({\a}-1)[L(L+1)-3(L+2{\a}-{\m}+
    {\b})^{2}]-({\l}+{\m}-L-{\b}-2{\a})[L(L+1) \\ -3({\m}-2{\a})^{2}]
    -\frac{1}{2}(L-2{\m}+4{\a}+{\b})(2L+3)(L+1+3{\b})\} \\
    +\sum_{k=1}^{\a}({\l}+2{\m}+3-4k)[({\l}+2{\m}+3-4k)^{2}+3-\frac{4}{3}
    L(L+1)-{\l}({\l}+2)] \end{array} $ \\
    &  \\
    1 &
    $ ({\l}+{\m}-L-{\b}-2{\a})({\m}-2{\a}-{\b})({\m}-2{\a}-{\b}-1)
    (L+{\l}+{\m}+2{\a}+{\b}+2) $ \\
    &  \\
     $-1$
 &
     $ -4{\a}({\a}-1)(L+2{\a}-{\m})(L+2{\a}-{\m}-1) $ \\
    \vspace{1pt} \\
    \hline
    \hline
    \end{tabular}
    \end{center}
     }
    \label{tab:matel}
    \end{table}
\ \ \ \ \
    \newpage
    \begin{table}
\caption{The parameters of the fits of the
energy levels and the transition ratios
    (Eqs (\protect\ref{eq:rat}) and (\protect\ref{eq:intra})) of the
 nuclei investigated are listed for the $(\l ,\m )$ multiplets which
    provide the best model descriptions.  The Hamiltonian parameters $g_{1}$,
$g_{2}$ and $g_{3}$ (Eq. (5)) are given in keV.  The quantities $\sigma_{E}$ 
(in keV) and $\sigma_{B}$ (dimensionless)
represent the energy (Eq. (\protect\ref{eq:se})) and the
transition (Eq. (\protect\ref{eq:sb})) rms factors respectively.  The splitting
    ratios $\Delta E_{2}$ (Eq. (\protect\ref{eq:split}),
dimensionless) and the
    vector--boson numbers $N$ (Eq. (\protect\ref{eq:NT})) are also
    given.}
    \bigskip
    \begin{center}
    \begin{tabular}{ccccccccc}
    \rule{0em}{2.2ex}
 & & & & & & & & \\
\hline\hline
Nucl&$\Delta E_{2}
$&$\l ,\m$&$\sigma_{E}$&$\sigma_{B}$&$g_{1}$&$g_{2}$&$g_{3}$&$N$ \\
\hline\hline
${}^{164}\rm Dy
$&$9.4$&$16, 2$&$14.1$&$0.52$&$-1.159$&$-0.321$&$-0.590$&$20$  \\
${}^{164}\rm Er
$&$8.4$&$18, 2$&$8.1$&$0.14$&$3.625$&$-0.238$&$-0.513$&$22$  \\
${}^{166}\rm Er
$&$8.8$&$16, 2$&$5.8$&$0.47$&$2.942$&$-0.235$&$-0.572$&$20$  \\
${}^{168}\rm Er
$&$9.3$&$20, 2$&$3.2$&$0.28$&$4.000$&$-0.181$&$-0.401$&$24$  \\
${}^{168}\rm Yb
$&$10.2$&$20, 2$&$7.9$&$0.27$&$0.500$&$-0.271$&$-0.501$&$24$  \\
${}^{172}\rm Yb
$&$17.6$&$\geq 80, 2$&$6.8$&$0.12$&$9.875$&$-0.017$&$-0.052$&$84$  \\
${}^{176}\rm Hf
$&$14.2$&$\geq 70, 2$&$15.0$&$0.17$&$9.547$&$-0.030$&$-0.062$&$74$  \\
${}^{178}\rm Hf
$&$11.6$&$34, 2$&$7.0$&$0.86$&$8.322$&$-0.083$&$-0.213$&$38$  \\
${}^{238}\rm U
$&$22.6$&$\geq 60, 2$&$1.6$&$0.08$&$-37.697$&$-0.360$&$-0.098$&$64$  \\
    \hline\hline
    \end{tabular}
    \end{center}
    \label{tab:favor}
    \end{table}

\ \ \ \ \ \
    \begin{table}
    \caption{Theoretical and experimental energy levels and
    transition ratios (Eqs (\protect\ref{eq:rat}) and
    (\protect\ref{eq:intra})) for the nuclei $^{164}$Dy, 
$^{164-168}$Er and
    $^{168}$Yb.  The corresponding $(\l ,\mu )$--values are also 
given.  The
    experimental data (used in the fits) for the energy levels are 
taken from
    \protect\cite{Sakai}, while the data for the E2 transitions are 
from
    \protect\cite{164Dy1,164Dy2,NDS164} (for $^{164}$Dy);
    \protect\cite{NDS164,164Er1} (for $^{164}$Er);
    \protect\cite{166Er1,166Er2,NDS166} (for $^{166}$Er);
    \protect\cite{168Er1,168Er2,NDS168} (for $^{168}$Er);
    \protect\cite{NDS168,168Yb1} (for $^{168}$Yb).
    The numbers in brackets refer to the uncertainties in the last 
digits of
    the experimental ratios.}

    \bigskip
    {\scriptsize
    \begin{center}
    \begin{tabular}{ccccccccccccc}
    \rule{0em}{2.2ex}
& & & & & & & & & & & & \\
\hline\hline
& & & & & &Nucleus&$(\l ,\m)$ & & & & & \\
\hline
$L$&$E_{g}^{Th}$&$E_{g}^{Exp}$&$E_{\gamma}^{Th}$&$E_{\gamma}^{Exp}$&
$R_{1}^{Th}$&$R_{1}^{Exp}$&$R_{2}^{Th}$&$R_{2}^{Exp}$&
$R_{3}^{Th}$&$R_{3}^{Exp}$&$R_{4}^{Th}$&$R_{4}^{Exp}$ \\
\hline\hline
& & & & & &${}^{164}\rm Dy$&(16,2)& & & & & \\
\hline
2&71.2&73.4&773.7&761.8&2.13&2.08(40)&0.11&0.082&--&--&--&-- \\
3&--&--&837.9&828.2&--&--&--&--&0.95&0.62&--&-- \\
4&237.2&242.2&924.3&961.0&8.73&9.10&0.26&0.26&--&--&1.39&1.30(17) \\
5&-- &-- &1030.5&1024.6&--&--&--&--&2.31&0.83&--&-- \\
6&496.1&501.3&1162.6&1154.0&31.54&--&0.45&--&--&--&1.05&1.14(28) \\
7&--&--&1309.6&--&--&--&--&--&4.91&--&--&-- \\
8&845.3&843.7&1492.5&--&407.5&--&0.66&--&--&--&0.97&0.97(34) \\
\hline
& & & & & &${}^{164}\rm Er$&(18,2)& & & & & \\
\hline
2&86.5&91.4&868.9&860.3&1.88&2.04(31)&0.088&0.11(5)&--&--&--&-- \\
3&--&--&949.2&946.3&--&--&--&--&0.74&0.89(7)&--&-- \\
4&288.0&299.5&1056.6&1058.3&5.92&--&0.20&--&--&--&1.40&1.18(33) \\
5&-- &-- &1190.2&1197.5&--&--&--&--&1.52&1.43(13)&--&-- \\
6&604.1&614.4&1352.1&1358.8&12.69&--&0.33&--&--&--&1.06&-- \\
7&--&--&1538.3&1545.1&--&--&--&--&2.65&--&--&-- \\
8&1033.8&1024.6&1756.5&1744.6&29.59&--&0.48&--&--&--&0.98&-- \\
\hline
& & & & & &${}^{166}\rm Er$&(16,2)& & & & & \\
\hline
2&76.8&80.6&790.9&785.9&1.83&1.86(10)&0.08&0.097(8)&--&--&--&-- \\
3&--&--&860.8&859.3&--&--&--&--&0.70&0.72(6)&--&-- \\

4&255.8&265.0&954.2&956.2&5.47&5.72(47)&0.20&0.26(7)&--&--&1.39&1.45(30) \\
5&-- &-- &1070.6&1075.3&--&--&--&--&1.41&1.43(15)&--&-- \\

6&536.6&545.4&1211.4&1215.9&10.72&12.25(75)&0.32&0.28&--&--&1.05&1.12(65) \\
7&--&--&1373.6&1376.0&--&--&--&--&2.36&--&--&-- \\

8&918.4&911.2&1563.0&1557.7&21.28&20.9(45)&0.48&--&--&--&0.96&1.05(95) \\
\hline
& & & & & &${}^{168}\rm Er$&(20,2) & & & & & \\
\hline
2&77.7&79.8&823.7&821.2&1.81&1.78(18)&0.082&0.066(16)&--&--&--&-- \\
3&--&--&896.5&895.8&--&--&--&--&0.68&0.62(6)&--&-- \\

4&258.8&264.1&993.8&994.7&5.34&4.81(78)&0.18&0.078(20)&--&--&1.40&1.53(18) \\
5&-- &-- &1115.1&1117.6&--&--&--&--&1.34&1.02(20)&--&-- \\
6&543.1&548.7&1261.6&1263.9&10.31&10.6(20)&0.29&0.19(2)&--&--&1.06&-- 
\\
7&--&--&1430.9&1432.9&--&--&--&--&2.19&1.62(16)&--&-- \\
8&929.9&928.3&1627.5&1624.5&19.94&--&0.42&--&--&--&0.99&-- \\
\hline
& & & & & &${}^{168}\rm Yb$&(20,2) & & & & & \\
\hline
2&82.3&87.7&990.0&983.8&1.97&2.06(36)&0.81&0.67(19)&--&--&--&-- \\
3&--&--&1066.3&1066.9&--&--&--&--&0.80&--&--&-- \\

4&273.9&286.6&1168.4&1171.2&6.82&6.72(135)&1.76&1.18(40)&--&--&1.40&-- \\
5&-- &-- &1295.1&1302.3&--&--&--&--&1.75&--&--&-- \\
6&574.1&585.3&1449.6&1445.1&17.26&17.3(42)&0.36&--&--&--&1.07&-- \\
7&--&--&1624.7&--&--&--&--&--&3.23&--&--&-- \\
8&981.5&970.1&1833.9&--&57.15&--&0.52&--&--&--&1.00&-- \\
\hline\hline

    \end{tabular}
    \end{center}
 }
\end{table}
\newpage
    \begin{table}
    \caption{The same as Table~3 but for the nuclei
    $^{172}$Yb, $^{176}$Hf, $^{178}$Hf, $^{238}$U.
    The experimental data for the energy levels are taken from
    \protect\cite{Sakai}, while the data for the E2 transitions are 
from
    \protect\cite{172Yb1,172Yb2,NDS172} (for $^{172}$Yb);
    \protect\cite{176Hf1,NDS176} (for $^{176}$Hf);
    \protect\cite{178Hf1,NDS178} (for $^{178}$Hf);
    \protect\cite{NDS238} (for $^{238}$U).}

    \bigskip
    {\scriptsize
    \begin{center}
    \begin{tabular}{ccccccccccccc}
    \rule{0em}{2.2ex}
 & & & & & & & & & & & & \\
\hline\hline
& & & & & &Nucleus&$(\l,\m)$ & & & & & \\
\hline
$L$&$E_{g}^{Th}$&$E_{g}^{Exp}$&$E_{\gamma}^{Th}$&$E_{\gamma}^{Exp}$&
$R_{1}^{Th}$&$R_{1}^{Exp}$&$R_{2}^{Th}$&$R_{2}^{Exp}$&
$R_{3}^{Th}$&$R_{3}^{Exp}$&$R_{4}^{Th}$&$R_{4}^{Exp}$ \\
\hline\hline
& & & & & &${}^{172}\rm Yb$&$(\geq80, 2)$ & & & & & \\
\hline
2&76.5&78.7&1480.4&1465.7&1.50&1.60(22)&0.05&0.105(13)&--&--&--&-- \\
3&--&--&1556.2&1549.2&--&--&--&--&0.43&0.50(6)&--&-- \\
4&256.0&260.1&1657.4&1657.9&3.18&3.12(48)&0.10&--&--&--&1.43&1.61(31) 
\\
5&-- &-- &1783.8&1792.3&--&--&--&--&0.65&--&--&-- \\
6&535.5&539.8&1933.3&--&4.22&--&0.13&--&--&--&1.10&0.98(41) \\
7&--&--&2111.3&--&--&--&--&--&0.79&--&--&-- \\
8&917.9&911.3&2314.8&--&4.97&--&0.15&--&--&--&1.04&1.20(47) \\
\hline
& & & & & &${}^{176}\rm Hf$&$(\geq70, 2)$ & & & & & \\
\hline
2&84.2&88.4&1361.4&1341.3&1.54&1.28(21)&0.06&0.13(5)&--&--&--&-- \\
3&--&--&1444.9&1445.8&--&--&--&--&0.48&0.61(23)&--&-- \\
4&280.8&290.2&1556.2&1540.2&3.56&--&0.11&--&--&--&1.43&-- \\
5&-- &-- &1695.3&1727.7&--&--&--&--&0.76&--&--&-- \\
6&589.6&597.0&1862.3&1861.9&5.04&--&0.15&--&--&--&1.10&-- \\
7&--&--&2047.6&--&--&--&--&--&0.99&--&--&-- \\
8&1010.7&998.0&2270.0&--&6.40&--&0.19&--&--&--&1.04&-- \\
\hline
& & & & & &${}^{178}\rm Hf$&(34,2) & & & & & \\
\hline
2&88.9&93.2&1180.2&1174.8&1.60&1.63(22)&0.06&0.11(6)&--&--&--&-- \\
3&--&--&1266.6&1268.9&--&--&--&--&0.51&0.46(8)&--&-- \\
4&296.4&306.6&1381.8&1384.6&3.84&5.9(10)&0.13&0.29(8)&--&--&1.42&-- 
\\
5&-- &-- &1525.7&1533.6&--&--&--&--&0.86&0.66(26)&--&-- \\
6&622.5&632.2&1698.5&1691.4&5.71&4.76(210)&0.18&--&--&--&1.09&-- \\
7&--&--&1899.0&--&--&--&--&--&1.16&--&--&-- \\
8&1067.0&1058.6&2129.2&--&7.61&--&0.23&--&--&--&1.03&-- \\
\hline
& & & & & &${}^{238}\rm U$&$(\geq60, 2)$ & & & & & \\
\hline
2&44.9&44.9&1062.2&1060.3&5.83&--&0.24&--&--&--&--&-- \\
3&--&--&1105.9&1105.7&--&--&--&--&4.57&--&--&-- \\
4&148.6&148.4&1165.9&1168.0&92.66&--&0.69&--&--&--&1.43&-- \\
5&--&--&1235.2&--&--&--&--&--&91.86&--&--&-- \\
6&308.1&307.2&1329.9&--&5.14&--&1.25&--&--&--&1.10&-- \\
7&--&--&1425.2&--&--&--&--&--&114.5&--&--&-- \\
8&519.4&518.3&1563.1&--&1.83&--&1.80&--&--&--&1.04&-- \\
9&--&--&1673.8&--&--&--&--&--&19.41&--&--&-- \\
10&777.1&775.7&1868.3&--&1.00&--&2.25&--&--&--&1.02&1.17(110) \\
12&1076.5&1076.5&--&--&--&--&--&--&--&--&1.01&1.11(125) \\
14&1413.4&1415.3&--&--&--&--&--&--&--&--&1.00&0.93(118) \\
16&1785.9&1788.2&--&--&--&--&--&--&--&--&1.00&1.00(68) \\
18&2193.9&2190.7&--&--&--&--&--&--&--&--&1.00&0.98(65) \\
    \hline\hline
    \end{tabular}
    \end{center}
 }
    \end{table}
\ \ \ \ \
\newpage
    \begin{center}
    {\bf Figure Captions}
    \end{center}
    \bigskip
{\bf Figure 1.} The energy rms factor $\sigma_{E}$ (Eq. (44)), obtained for the
    nucleus $^{168}$Er, is plotted as a function of the quantum number $\l$ at
    $\m =2$ (circlets), $\m =4$ (squares), $\m =6$ (triangles), and
    $\m =8$ (asterisks).
    \smallskip \\
    {\bf Figure 2.} The Hamiltonian parameters $g_{1}$, $g_{2}$,
    $g_{3}$ (Eq. (5)) and the ratio $g_{3}/g_{2}$, adjusted for the nucleus
    $^{168}$Er,  are plotted (in parts (a), (b), (c), and (d) respectively)
as functions of the quantum number $\l$ at
    $\m =2$ (circlets), $\m =4$ (squares), $\m =6$ (triangles), and
    $\m =8$ (asterisks).
    \smallskip \\
    {\bf Figure 3.} The same as Fig. 1 but for the nucleus $^{166}$Er.
    \smallskip \\
    {\bf Figure 4.} The same as Fig. 1 but for the nucleus $^{164}$Er.
    \smallskip \\
{\bf Figure 5.} The energy rms factor $\sigma_{E}$ (Eq. (44)), obtained for the
    nuclei $^{164}$Dy and $^{168}$Yb (shown in parts (a) and (b) 
respectively), is plotted as a function of the quantum number $\l$ at $\m =2$.
    \smallskip \\
{\bf Figure 6.} The energy rms factor $\sigma_{E}$ (Eq. (44)), obtained for the
    nucleus $^{172}$Yb, is plotted as a function of the quantum number $\l$ at
    $\m =2$ (circlets), $\m =4$ (squares), and $\m =6$ (triangles).
    \smallskip \\
    {\bf Figure 7.} The same as Fig. 5 but for $^{176}$Hf (part (a)) and
    $^{238}$U (part (b)).
    \smallskip \\
    {\bf Figure 8.} The same as Fig. 5 but for the nucleus $^{178}$Hf.

    \end{document}